\documentclass[12pt]{spieman}  
\usepackage{amsmath,amsfonts,amssymb}
\usepackage{graphicx}
\usepackage{setspace}
\usepackage{tocloft}
\usepackage{lineno}
\usepackage{array}
\usepackage{multirow}

\title{SUPPPPRESS: Prototyping and testing liquid-crystal vector vortex coronagraphs with reduced polarization leakage}

\author[a, *, +]{Rico Landman}
\author[b, *]{David Doelman}
\author[b]{Jeroen Rietjens}
\author[a, c]{Iva Laginja}
\author[c]{Pierre Baudoz}
\author[d]{Kristien Peeters}
\author[d]{Chris van Dijk}
\author[e]{Yuki Nishie}
\author[e]{Yuta Watanabe}
\author[b]{Alexander Eigenraam}
\author[b]{Mario Vretenar}
\author[f]{Joost van den Born}
\author[c]{Rapha\"el Galicher}
\author[c]{Johan Mazoyer}
\author[c]{Axel Potier}
\author[g]{Mariya Krasteva}
\author[g]{Matteo Taccola}
\author[a]{Felix Bettonvil}
\author[a]{Frans Snik}

\affil[*]{These authors contributed equally to this work.}
\affil[a]{NOVA, Einsteinweg 55,
2333 CC Leiden, The Netherlands}
\affil[b]{SRON, Space Research Organisation Netherlands, Niels Bohrweg 4, 2333 CA Leiden, The Netherlands}
\affil[c]{LIRA, Observatoire de Paris, Universit\'e PSL, CNRS, Universit\'e Paris Cit\'e, Sorbonne Universit\'e, CY Cergy Paris Universit\'e, 92190 Meudon, France}
\affil[d]{cosine Remote Sensing B.V., Warmonderweg 14, 2171 AH Sassenheim, The Netherlands}
\affil[e]{ColorLink Japan, Ltd., 1-5-5 Minamihon-cho, Joetsu-shi, Niigata, 943-0841, Japan}
\affil[f]{NOVA Optical Infrared Instrumentation Group at ASTRON, Oude Hoogeveensedijk 4, 7991 PD Dwingeloo, The Netherlands}
\affil[g]{European Space Agency, ESTEC, Keplerlaan 1, 2200 AG Noordwijk, the Netherlands}

\cftpagenumbersoff{figure}
\cftpagenumbersoff{table} 
\RequirePackage[normalem]{ulem} 
\RequirePackage{color}\definecolor{RED}{rgb}{1,0,0}\definecolor{BLUE}{rgb}{0,0,1} 
\providecommand{\DIFadd}[1]{{#1}} 
\providecommand{\DIFdel}[1]{}                      
\providecommand{\DIFaddbegin}{} 
\providecommand{\DIFaddend}{} 
\providecommand{\DIFdelbegin}{} 
\providecommand{\DIFdelend}{} 
\providecommand{\DIFaddFL}[1]{\DIFadd{#1}} 
\providecommand{\DIFdelFL}[1]{\DIFdel{#1}} 
\providecommand{\DIFaddbeginFL}{} 
\providecommand{\DIFaddendFL}{} 
\providecommand{\DIFdelbeginFL}{} 
\providecommand{\DIFdelendFL}{} 
\newcommand{\DIFscaledelfig}{0.5}
\RequirePackage{settobox} 
\RequirePackage{letltxmacro} 
\newsavebox{\DIFdelgraphicsbox} 
\newlength{\DIFdelgraphicswidth} 
\newlength{\DIFdelgraphicsheight} 
\LetLtxMacro{\DIFOincludegraphics}{\includegraphics} 
\newcommand{\DIFaddincludegraphics}[2][]{{{\DIFOincludegraphics[#1]{#2}}}} 
\newcommand{\DIFdelincludegraphics}[2][]{
\sbox{\DIFdelgraphicsbox}{\DIFOincludegraphics[#1]{#2}}
\settoboxwidth{\DIFdelgraphicswidth}{\DIFdelgraphicsbox} 
\settoboxtotalheight{\DIFdelgraphicsheight}{\DIFdelgraphicsbox} 
\scalebox{\DIFscaledelfig}{
\parbox[b]{\DIFdelgraphicswidth}{\usebox{\DIFdelgraphicsbox}\\[-\baselineskip] \rule{\DIFdelgraphicswidth}{0em}}\llap{\resizebox{\DIFdelgraphicswidth}{\DIFdelgraphicsheight}{
\setlength{\unitlength}{\DIFdelgraphicswidth}
\begin{picture}(1,1)
\thicklines\linethickness{2pt} 
{\color[rgb]{1,0,0}\put(0,0){\framebox(1,1){}}}
{\color[rgb]{1,0,0}\put(0,0){\line( 1,1){1}}}
{\color[rgb]{1,0,0}\put(0,1){\line(1,-1){1}}}
\end{picture}
}\hspace*{3pt}}} 
} 
\LetLtxMacro{\DIFOaddbegin}{\DIFaddbegin} 
\LetLtxMacro{\DIFOaddend}{\DIFaddend} 
\LetLtxMacro{\DIFOdelbegin}{\DIFdelbegin} 
\LetLtxMacro{\DIFOdelend}{\DIFdelend} 
\DeclareRobustCommand{\DIFaddbegin}{\DIFOaddbegin \let\includegraphics\DIFaddincludegraphics} 
\DeclareRobustCommand{\DIFaddend}{\DIFOaddend \let\includegraphics\DIFOincludegraphics} 
\DeclareRobustCommand{\DIFdelbegin}{\DIFOdelbegin \let\includegraphics\DIFdelincludegraphics} 
\DeclareRobustCommand{\DIFdelend}{\DIFOaddend \let\includegraphics\DIFOincludegraphics} 
\LetLtxMacro{\DIFOaddbeginFL}{\DIFaddbeginFL} 
\LetLtxMacro{\DIFOaddendFL}{\DIFaddendFL} 
\LetLtxMacro{\DIFOdelbeginFL}{\DIFdelbeginFL} 
\LetLtxMacro{\DIFOdelendFL}{\DIFdelendFL} 
\DeclareRobustCommand{\DIFaddbeginFL}{\DIFOaddbeginFL \let\includegraphics\DIFaddincludegraphics} 
\DeclareRobustCommand{\DIFaddendFL}{\DIFOaddendFL \let\includegraphics\DIFOincludegraphics} 
\DeclareRobustCommand{\DIFdelbeginFL}{\DIFOdelbeginFL \let\includegraphics\DIFdelincludegraphics} 
\DeclareRobustCommand{\DIFdelendFL}{\DIFOaddendFL \let\includegraphics\DIFOincludegraphics} 
\RequirePackage{listings} 
\RequirePackage{color} 
\lstdefinelanguage{DIFcode}{ 
  moredelim=[il][\color{red}\sout]{\%DIF\ <\ }, 
  moredelim=[il][\color{blue}\uwave]{\%DIF\ >\ } 
} 
\lstdefinestyle{DIFverbatimstyle}{ 
	language=DIFcode, 
	basicstyle=\ttfamily, 
	columns=fullflexible, 
	keepspaces=true 
} 
\lstnewenvironment{DIFverbatim}{\lstset{style=DIFverbatimstyle}}{} 
\lstnewenvironment{DIFverbatim*}{\lstset{style=DIFverbatimstyle,showspaces=true}}{} 

\begin{document} 
\maketitle

\begin{abstract}
The vortex coronagraph is one of the most promising candidates for the Habitable Worlds Observatory (HWO) due to its excellent theoretical performance for an off-axis telescope. A practical realization can be achieved using liquid-crystal polymers to form a vector vortex coronagraph (VVC). Reaching the $10^{-10}$ contrast required for Earth-like planet detection is, however, limited by polarization leakage caused by wavelength-dependent deviations from half-wave retardance. This effect can be mitigated using multi-layer twisted retarders to minimize leakage, and by combining the VVC with multiple polarization gratings (mgVVC) to diffract the polarization leakage out of the science path. We present recent progress within the ESA-funded SUPPPPRESS project, which aims to advance the manufacturing, assembly, and testing of high-performance VVCs. Central singularities of 2 and 6 $\mu m$ have been achieved for charge 2 and charge 6 VVCs, respectively, with patterning accuracies better than 1 degree root-mean-square error. Fabrication procedures have been developed to produce individual components with a polarization leakage of $3\times 10^{-4}$ over a 10\% bandwidth and $8\times 10^{-4}$ over a 20$\%$ bandwidth. We also report on the development of assembly and alignment procedures \DIFdelbegin \DIFdel{of }\DIFdelend \DIFaddbegin \DIFadd{for }\DIFaddend mgVVCs and their metrology. Furthermore, we present initial high-contrast tests at the THD2 bench for both regular VVCs and a double-grating VVC. The \DIFdelbegin \DIFdel{double grating }\DIFdelend \DIFaddbegin \DIFadd{double-grating }\DIFaddend VVC reaches an average contrast between 3 and 10 $\lambda/D$ of $2\times 10^{-8}$ over a small bandwidth and $6 \times 10^{-8}$ over a 10\% bandwidth. Finally, we report on successful space-environment tests of the assembled liquid-crystal masks.

\end{abstract}


{\noindent \footnotesize\textbf{+} \linkable{rlandman@strw.leidenuniv.nl} }


{}

\section{Introduction}
The direct imaging and characterization of rocky planets in the habitable zones of nearby stars is one of the most ambitious goals in modern astrophysics. Achieving these goals requires the suppression of starlight by factors of $10^{10}$, allowing the faint reflected light from an exoplanet to be detected. The upcoming Habitable Worlds Observatory (HWO)\cite{2026arXiv260111803F}  is designed to meet this challenge by combining a large, diffraction-limited space telescope with ultra-stable wavefront control and high-performance coronagraphy. In addition to detecting these faint planetary companions, spectral and polarimetric measurements will enable the identification of key atmospheric molecular features, such as oxygen and water vapor, and may even reveal surface oceans through their polarization signatures \cite{2023MNRAS.524.5477V}. Consequently, both broadband performance and polarization control are essential design drivers for future space-based coronagraphs.

Among the various coronagraph concepts under development, the (apodized) vortex coronagraph \cite{Foo2005O_vortex} is one of the \DIFdelbegin \DIFdel{the }\DIFdelend most promising candidates for the HWO mission \cite{ Ruane2018JATIS...4a5004R, 2024JATIS..10c5004M, 2024SPIE13092E..66B}. It offers an attractive balance between technological maturity, optical simplicity, and high throughput, particularly for off-axis telescope architectures. Compared to the classical Lyot coronagraph, the vortex coronagraph provides increased throughput and robustness to low-order aberrations, while maintaining small inner working angles \cite{Mawet2010SPIE_VVC_sensitivity, 2024JATIS..10c5004M}. Its operating principle is based on introducing an azimuthal phase ramp in the focal plane that converts the incoming starlight into an optical vortex. This redistributes the stellar light outside the pupil by destructive interference, where it can be blocked by a downstream Lyot stop, while allowing off-axis planetary light to pass largely unaffected.

Over the past two decades, several optical implementations of the vortex coronagraph have been proposed and experimentally demonstrated. The majority of these designs rely on the geometric (Pancharatnam–Berry) phase \cite{pancharatham, 1987JMOp...34.1401B} to realize a vector vortex coronagraph (VVC), where the local orientation of a half-wave retarder imposes the desired azimuthal phase ramp. Such devices can, for example, be manufactured using patterned liquid-crystal polymers (LCPs) \cite{Mawet2009OExpr_VVC_LCP, Murakami2013OExpr_achromatic_vvc, 2019JOSAB..36D..13S}, photonic crystals \cite{Murakami2012SPIE}, or subwavelength gratings \cite{mawet2005ApJ_agpm, 2013A&A...553A..98D,2016A&A...595A.127V}. Vector vortex coronagraphs have been deployed on various ground-based instruments  \cite{Serabyn2010Natur.464.1018S,Mawet2010ApJ_VVC_palomar, 2018PASP..130c5001K,2018AJ....156..156X, 2024SPIE13097E..15O}, and deep contrasts have been demonstrated in both monochromatic and broadband light on space-oriented testbeds \DIFdelbegin \DIFdel{\mbox{
\cite{2019JOSAB..36D..13S,2020AJ....159...79L, 2022SPIE12180E..24R, 2020A&A...635A..11G, 2023SPIE12680E..2CD, 2025JATIS..11c9001M}}\hskip0pt
}\DIFdelend \DIFaddbegin \DIFadd{\mbox{
\cite{2019JOSAB..36D..13S,2020AJ....159...79L, 2022SPIE12180E..24R, Galicher2020AFamilyofPhaseMasks, 2024SPIE13092E..1YL,2023SPIE12680E..2CD, 2025JATIS..11c9001M}}\hskip0pt
}\DIFaddend .

One of the main performance limitations for these VVCs is polarization leakage caused by deviations of the retardance from half-wave, especially when operated over a larger spectral bandwidth \cite{Mawet2009OExpr_VVC_LCP}. This leakage results in residual starlight that can significantly limit the achievable contrast. To combat this, the VVC is usually operated with polarization filtering, where it is sandwiched between linear polarizers and quarter-wave plates to remove the polarization leakage. However, this reduces the planet throughput by \DIFdelbegin \DIFdel{a factor }\DIFdelend \DIFaddbegin \DIFadd{at least a factor of }\DIFaddend two. Even with polarization filtering, reaching 10$^{-10}$ is still a challenge due to the strict requirements on the polarization leakage and its filtering. In recent years, polarization-independent scalar vortex coronagraphs have also been proposed and demonstrated \cite{2013MNRAS.435..565E, 2019SPIE11117E..1FR, 2023JATIS...9b5001D, 2022SPIE12180E..5HD, 2024SPIE13092E..21D, 2023SPIE12680E..0QK, 2025JATIS..11b5002K}. While these scalar designs eliminate polarization leakage, they instead suffer from step functions in the phase pattern and strong chromatic variations in the effective phase ramp, which limit their broadband performance. To address this, ongoing developments are exploring the use of metasurfaces \cite{2023SPIE12680E..0QK, 2023SPIE12680E..0PP,2025JATIS..11b5002K} and modified phase patterns \cite{2023JATIS...9b5001D, 2024JATIS..10a5001D, 10.1117/12.3064060} to mitigate the chromaticity.

Several approaches have also been developed to decrease the polarization leakage of the VVC. First, increasing the degrees of freedom of the retarder by using multiple layers can help improve the broadband retardance performance \cite{Mawet2010SPIE_VVC_sensitivity, 2013OExpr..21..404K}. More recently, the use of polarization gratings has been suggested as a way to reduce polarization leakage \cite{2020PASP..132d5002D}. Such a multi-grating vector vortex coronagraph (mgVVC) combines the vortex mask with one or more polarization gratings that angularly separates the polarization leakage from the main beam. By appropriately recombining the diffracted beams, the unwanted leakage can be redirected away from the science path, effectively suppressing its impact on contrast. \DIFdelbegin \DIFdel{A }\DIFdelend \DIFaddbegin \DIFadd{We obtained a }\DIFaddend first prototype of such a multi-grating VVC\DIFdelbegin \DIFdel{was }\DIFdelend \DIFaddbegin \DIFadd{, }\DIFaddend manufactured by ImagineOptix\DIFdelbegin \DIFdel{and tested by }\DIFdelend \DIFaddbegin \DIFadd{, in 2021, and initial lab tests were shown in }\DIFaddend Doelman et al.~\DIFdelbegin \DIFdel{\mbox{
\cite{2023SPIE12680E..2CD}}\hskip0pt
, demonstrating the feasibility of the concept. The measurements confirmed that the additional gratings effectively removed most of the polarization leakage, but }\DIFdelend \DIFaddbegin \DIFadd{2022~\mbox{
\cite{2023SPIE12680E..2CD}}\hskip0pt
. However, }\DIFaddend the final performance was \DIFaddbegin \DIFadd{likely }\DIFaddend limited by imperfections in the liquid-crystal manufacturing. \DIFaddbegin \DIFadd{Additionally, this prototype targeted only monochromatic performance at visible wavelengths. }\DIFaddend Continued development is therefore required to improve the patterning accuracy, minimize \DIFaddbegin \DIFadd{broadband }\DIFaddend polarization leakage, and further mature the concept of mgVVCs. 

We are now working together with \DIFdelbegin \DIFdel{Colorlink }\DIFdelend \DIFaddbegin \DIFadd{ColorLink }\DIFaddend Japan, Ltd.\footnote{https://www.colorlink.co.jp/en/}~to improve manufacturing capabilities in order to obtain better (mg)VVCs. Our technology provides the following unique benefits:
\begin{itemize}
    \item Direct-write allows for arbitrary phase patterns with small pixels, required for mgVVCs. \cite{2014OExpr..2212691M}
    \item Self-aligning multi-twist retarders to obtain high broadband diffraction efficiency\cite{2013OExpr..21..404K}.
\end{itemize}

With such broadband polarization-based coronagraphs we can enable the following opportunities on system-level:
\begin{itemize}
    \item Implementation of a single coronagraph that can be operated for a large range of spectral bandpasses.
    \item A large instantaneous wavelength coverage for spectroscopy.
    \item Integrated focal plane wavefront sensing.\cite{2012A&A...545A.151R,2021SPIE11823E..1TM}
    \item Integrated polarimetry.\cite{2014SPIE.9147E..7US, 2023SPIE12680E..0FM}
\end{itemize}

The SUPPPPRESS\cite{2024SPIE13092E..27L, 2025SPIE13699E..0HD} (Substantiating Unique Patterned Polarization-sensitive Polymer Photonics for Research of
Exoplanets with Space-based Systems) project addresses these opportunities and challenges by developing a new generation of low-leakage vector vortex coronagraphs for future space telescopes. The project is funded by the European Space Agency and brings together several European partners: NOVA, Leiden University, SRON, cosine \DIFaddbegin \DIFadd{Remote Sensing B.V.}\DIFaddend , and Paris Observatory. The main objectives are to improve the manufacturing quality of multi-layer liquid-crystal masks produced by ColorLink Japan, to establish robust assembly and metrology procedures for multi-grating devices, \DIFdelbegin \DIFdel{and }\DIFdelend to validate their optical performance on high-contrast testbeds\DIFaddbegin \DIFadd{, and to make strides in the space qualification of these components}\DIFaddend .  

In this paper, we present the first results of the SUPPPPRESS project. We describe recent progress in mask manufacturing at ColorLink Japan, the procedures developed for assembly of mgVVCs, and the metrology methods used to characterize the resulting optical elements. We also show experimental results obtained on the Tr\`es Haute Dynamique 2 (THD2) testbed for regular VVCs and a double-grating VVC (dgVVC).
\DIFdelbegin \DIFdel{A follow-up publication will report on the performance of triple-grating devices currently in fabrication and any other further performance upgrades.
}\DIFdelend

\section{Improving the liquid-crystal mask manufacturing}
\begin{figure}[htbp]
\centering
\includegraphics[width = \linewidth]{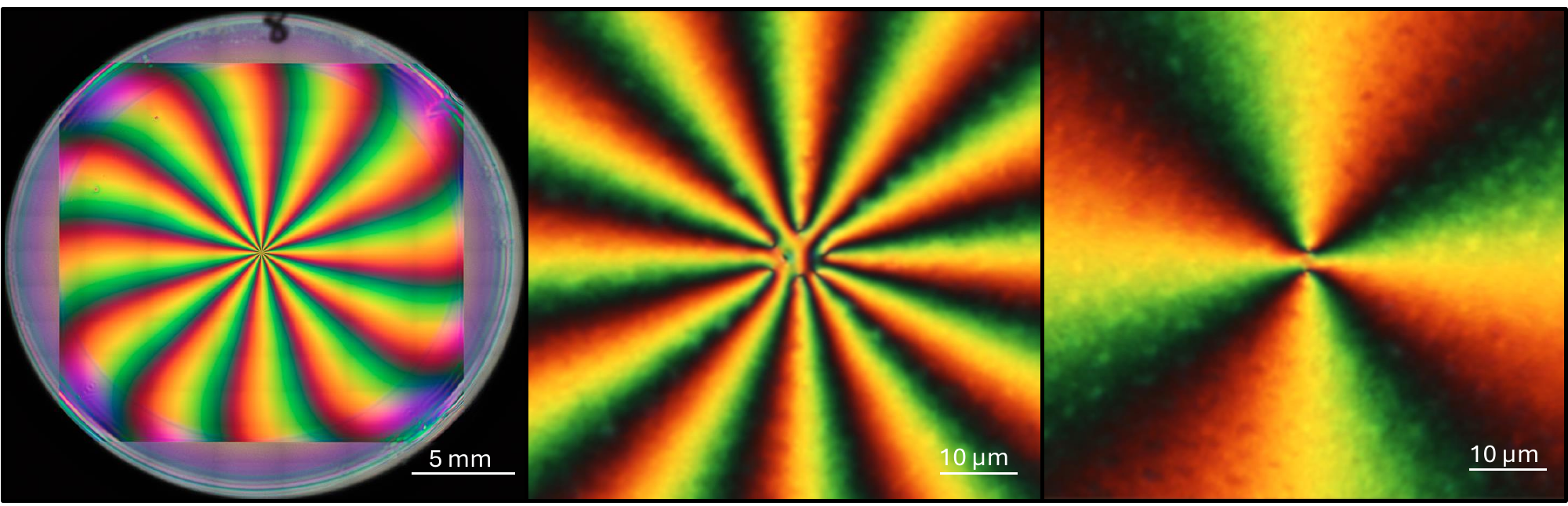}
\DIFaddbeginFL \caption{
\textbf{\DIFaddFL{Left:}} \DIFaddFL{Stitched microscope image between crossed polarizers of a manufactured charge 6 VVC (VVC6-1). The grid-like structure is the result of the stitching software and not a real feature. }\textbf{\DIFaddFL{Middle:}} \DIFaddFL{Microscope image between crossed polarizers of the center of the same charge 6 VVC. }\textbf{\DIFaddFL{Right:}} \DIFaddFL{Microscope image between crossed polarizers of the center of a manufactured charge 2 VVC (VVC2-1).
}}
\DIFaddendFL \label{fig:microscope-image} 
\DIFdelbeginFL 
{
\textbf{\DIFdelFL{Left:}} 
\DIFdelFL{Stitched microscope image between crossed polarizers of a manufactured 3TR charge 6 VVC. The grid-like structure is the result of the stitching software and not a real feature. }\textbf{\DIFdelFL{Middle:}} 
\DIFdelFL{Microscope image between crossed-polarizers of the center of a manufactured charge 6 VVC. }\textbf{\DIFdelFL{Right:}} 
\DIFdelFL{Microscope image between crossed-polarizers of the center of a manufactured charge 2 VVC.
}}
\DIFdelendFL \end{figure}

The VVCs used throughout this work \DIFdelbegin \DIFdel{are }\DIFdelend \DIFaddbegin \DIFadd{were }\DIFaddend fabricated using the direct-write liquid-crystal patterning technique \cite{2014OExpr..2212691M}. A photo-alignment layer is first deposited on a substrate, after which the direct-write system imprints the required fast-axis orientation pattern by exposing the layer with linearly polarized UV light. \DIFaddbegin \DIFadd{We used 1.1 mm thick fused silica substrates with an anti-reflection (AR) coating with $<0.2\%$ reflectivity between 600 and 800 nm and $<1\%$ at 355 nm on the non-LCP side. }\DIFaddend Subsequently, birefringent liquid-crystal layers are spin-coated and cured on top, which \DIFdelbegin \DIFdel{will }\DIFdelend orient along the imprinted local orientation of the photo-alignment layer. To obtain broadband efficiency, multiple self-aligning layers are then added in sequence, each with a carefully chosen thickness and twist angle to achieve an overall half-wave retardance across the desired wavelength range \cite{2013OExpr..21..404K}. \DIFdelbegin \DIFdel{The manufacturing of our masks is done by Colorlink Japan, Ltd. }\DIFdelend A stitched microscope image of \DIFdelbegin \DIFdel{an example manufactured }\DIFdelend \DIFaddbegin \DIFadd{a manufactured charge 6 }\DIFaddend VVC is shown \DIFdelbegin \DIFdel{in }\DIFdelend \DIFaddbegin \DIFadd{on the left panel of }\DIFaddend Fig.~\ref{fig:microscope-image}. \DIFaddbegin \DIFadd{An overview of the manufactured prototypes and assemblies that are used throughout this work is given in Table~\ref{tab:overview}. These consist of three main manufacturing runs: In the first batch, manufactured in August 2025, we produced the components for a dgVVC prototype. Next, in September 2025, we manufactured a number of VVCs of different charges. Finally, in October 2025, we manufactured a batch of components for a triple-grating VVC (tgVVC) prototype. Unfortunately, the assembly of this tgVVC failed. However, these components demonstrate the quality of the latest manufacturing batch and, therefore, we still show their component-level performance in this work. The space environment tests use a separate set of samples, including ones that exhibited manufacturing or assembly errors, or possessed poorer metrology results than those listed in Table~\ref{tab:overview}.}\\
\DIFaddend 

\DIFaddbegin \begin{table}[]
\footnotesize
\begin{tabular}{l||l|l|l}
\DIFaddFL{Prototype          }& \DIFaddFL{Description                                    }& \DIFaddFL{Manufacturing date }& \DIFaddFL{Figures      }\\ \hline\hline
\DIFaddFL{FG6-1              }& \DIFaddFL{Charge 6 forked grating with a 10 $\mu$m pitch }& \DIFaddFL{August 2025        }& \DIFaddFL{6, 7         }\\ \hline
\DIFaddFL{PG10-1             }& \DIFaddFL{Polarization grating with a 10 $\mu$m pitch    }& \DIFaddFL{August 2025        }& \DIFaddFL{6            }\\ \hline
\DIFaddFL{dgVVC              }& \DIFaddFL{Double-grating VVC (PG10-1 + FG6-1)         }& \DIFaddFL{August 2025        }& \DIFaddFL{6, 10, 11, 12, 14 }\\ \hline
\DIFaddFL{VVC-2, VVC-4,  }& \DIFaddFL{VVCs with different charges }& \DIFaddFL{September 2025}& \DIFaddFL{1, 3, 13 }\\\DIFaddFL{VVC-6-1, VVC6-2 }&                        &      &            \\ \hline
\DIFaddFL{PG10-2,10-3     }& \DIFaddFL{New batch of polarization gratings (10 $\mu$m)            }& \DIFaddFL{October 2025       }& \DIFaddFL{3, 5            }\\ \hline
\DIFaddFL{FG6-2              }& \DIFaddFL{Charge 6 forked grating with a 30 $\mu$m pitch }& \DIFaddFL{October 2025       }& \DIFaddFL{3, 5, 9        }\\
\end{tabular}
\vspace{5pt}
\caption{\DIFaddFL{Overview of the manufactured prototypes, their manufacturing date and the figures that these prototypes are used for. Abbreviations: FG: forked grating, PG: polarization grating, VVC: vector vortex coronagraph, dgVVC: double-grating VVC}}
\label{tab:overview}
\end{table}

\DIFaddend 

The main stellar leakage terms of VVCs are discussed in Ruane et al.\DIFaddbegin \DIFadd{~}\DIFaddend (2022) \cite{2022SPIE12180E..24R} and can be summarized as follows:
\begin{itemize}
    \item Localized defects, most importantly in the central singularity of the vortex.
    \item Imperfect vortex phase mask from fast axis orientation errors.
    \item Polarization leakage due to deviations from half-wave retardance.
\end{itemize}

In the following sections, we will discuss the quality of these aspects for the masks manufactured by \DIFdelbegin \DIFdel{Colorlink }\DIFdelend \DIFaddbegin \DIFadd{ColorLink }\DIFaddend Japan. We note that there is ongoing effort to improve the manufacturing further.

\subsection{Central singularity and defects}
Fig. ~\ref{fig:microscope-image} shows a \DIFdelbegin \DIFdel{zoomed }\DIFdelend \DIFaddbegin \DIFadd{microscope }\DIFaddend image of a manufactured charge-6 and charge-2 vortex coronagraph. These components have a central defect of \DIFdelbegin \DIFdel{about }\DIFdelend $\sim$6 $\mu m$ and $\sim$2 $\mu m$ in diameter for the charge 6 and charge 2 VVC respectively. This is slightly larger than the results presented by Nersisyan et al.~(2013)\cite{2013OExpr..21.8205N} and Serabyn et al.~(2019)\cite{2019JOSAB..36D..13S}\DIFdelbegin \DIFdel{. Still, this is more than a factor 10 smaller than $F\lambda$ }\DIFdelend \DIFaddbegin \DIFadd{, but smaller than the results presented by Roberts et al.~2022\mbox{
\cite{9843574}}\hskip0pt
. The stellar leakage due to this central singularity can be approximated to first order as $(d/F\lambda)^4$ \mbox{
\cite{2019JOSAB..36D..13S}}\hskip0pt
, where $d$ is the size of the central defect, $F$ is the F-number }\DIFaddend at the coronagraphic focal plane\DIFdelbegin \DIFdel{for the }\DIFdelend \DIFaddbegin \DIFadd{, and $\lambda$ is the wavelength. A defect of 6 $\mu m$ in diameter gives an on-axis leakage of about $3 \times 10 ^{-6}$ for the F/110 beam and a wavelength of 780 nm at the }\DIFaddend THD2\DIFdelbegin \DIFdel{bench \mbox{
\cite{2018SPIE10706E..2OB}}\hskip0pt
. }\DIFdelend \DIFaddbegin \DIFadd{, which we confirm with our diffraction simulation in Section~\ref{sec:performance_sim}. The impact of the central defect will likely be more significant at the desired F-number for an HWO coronagraph, depending on packaging constraints. }\DIFaddend An opaque dot covering the central defect may \DIFdelbegin \DIFdel{further }\DIFdelend reduce the impact of this central singularity \cite{Ruane2018JATIS...4a5004R}. In this work, we do not include this opaque dot on the vortices, but these are planned to be tested in the near future.

Additionally, defects can be found further away from the center, as for example shown by the two dots in the top left of Fig.~\ref{fig:microscope-image}. These are likely the result of contamination from particles, \DIFdelbegin \DIFdel{which are subsequently spin-coated on top of }\DIFdelend \DIFaddbegin \DIFadd{that are subsequently covered during spin-coating of the next layer}\DIFaddend . Initially, these defects were much more common, but steps were taken to minimize the chance of contamination. Smaller, sub-micron sized defects can occasionally also be seen throughout the optic, which are likely the result of crystallization of the LCP material. The occurrence of different types of defects can decrease the yield of the manufacturing. Further steps are being considered to improve the yield and quality control of the manufacturing.

\subsection{Fast axis errors}
To assess the patterning quality of the manufactured components, we perform measurements of the sample between crossed polarizers using a microscope \DIFaddbegin \DIFadd{with an RGB camera}\DIFaddend . Images are acquired every $5^\circ$ over a full $360^\circ$ rotation of the second polarizer \DIFaddbegin \DIFadd{and we only use the red channel in the remainder of this analysis}\DIFaddend . Because we observed significant beam wobble, the images were first aligned with one another. We then fit a differentiable Jones-matrix model of the spatially variable retarder to the processed data. The free parameters in this model are the fast axis at each pixel and the global retardance.
\DIFdelbegin \DIFdel{For optics composed of multiple twisted retarders, we additionally fit for the twist and thickness of each LCP layer.
}\DIFdelend 

\begin{figure}[htbp]
\centering
\includegraphics[width = \linewidth]{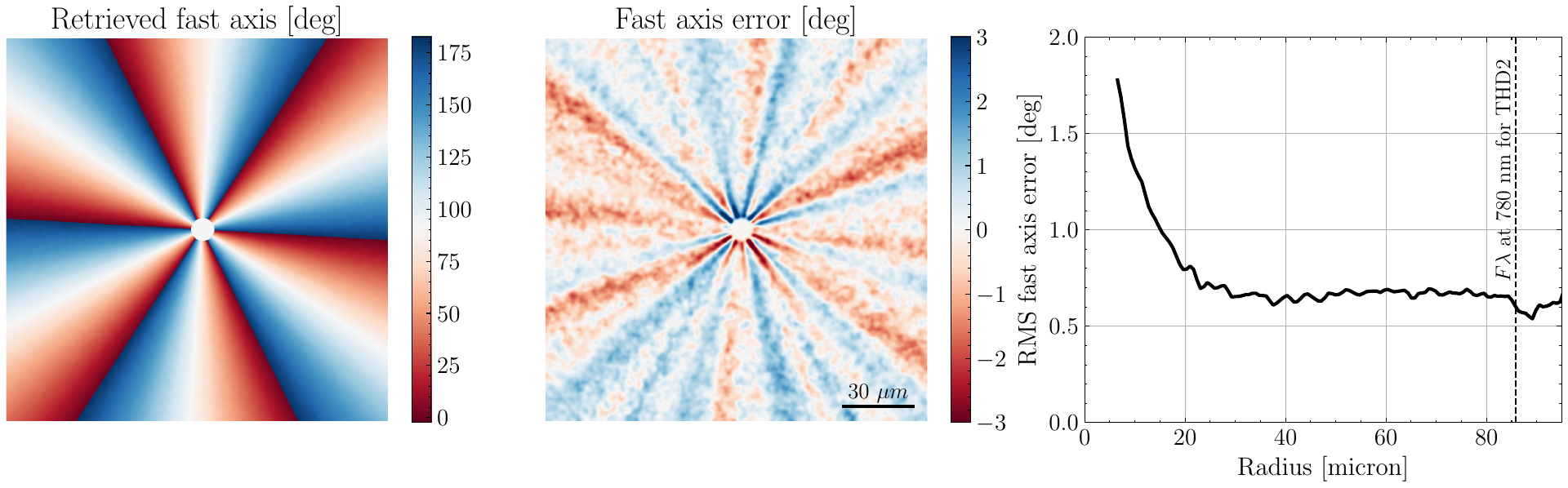}
\caption[1TR fast-axis writing result] 
{\label{fig:1tr-writing-result} 
Retrieved fast axis error for \DIFdelbeginFL \DIFdelFL{one of the manufactured }\DIFdelendFL \DIFaddbeginFL \DIFaddFL{a calibration }\DIFaddendFL charge 6 \DIFdelbeginFL \DIFdelFL{vortex coronagraphs}\DIFdelendFL \DIFaddbeginFL \DIFaddFL{VVC}\DIFaddendFL , along with the fast axis error compared to a perfect vortex. \textbf{Right:} Azimuthal root-mean-square (RMS) error of the fast axis as a function of distance from the center. Also indicated is the size of \DIFaddbeginFL \DIFaddFL{the }\DIFaddendFL PSF at the coronagraphic focal plane of the THD2 bench.
}
\end{figure}

Before manufacturing the \DIFdelbegin \DIFdel{real }\DIFdelend \DIFaddbegin \DIFadd{final }\DIFaddend components, we produced a series of calibration samples \DIFaddbegin \DIFadd{on Eagle XG glass with untwisted LCP layers}\DIFaddend . The retrieved fast-axis error from each sample was iteratively corrected in the subsequent one by modifying the voltage-response curve in the driving of the Pockels cell in the direct-write system. \DIFdelbegin \DIFdel{Figure}\DIFdelend \DIFaddbegin \DIFadd{Fig.}\DIFaddend ~\ref{fig:1tr-writing-result} shows the retrieved fast axis for \DIFdelbegin \DIFdel{one of the }\DIFdelend \DIFaddbegin \DIFadd{a }\DIFaddend calibration charge 6 \DIFdelbegin \DIFdel{VVCs}\DIFdelend \DIFaddbegin \DIFadd{VVC}\DIFaddend , along with the corresponding deviation from an ideal charge 6 design. We \DIFdelbegin \DIFdel{obtain an aximuthal }\DIFdelend \DIFaddbegin \DIFadd{show the result of the calibration pattern produced before VVC6-1 instead of the final sample as the data from the untwisted retarder is significantly easier to interpret. A second calibration produced right after VVC6-1 resulted in similar errors and we thus assume VVC6-1 exhibits similar errors as well. We obtain an azimuthal }\DIFaddend root-mean-square fast axis error of $0.7$ degrees \DIFaddbegin \DIFadd{or $0.012$ radians }\DIFaddend (1.4 degrees \DIFaddbegin \DIFadd{or $\sim \pi/130$ radians }\DIFaddend in phase) outside the central 20 $\mu$m. The \DIFdelbegin \DIFdel{overall patterning }\DIFdelend \DIFaddbegin \DIFadd{exact shape of the patterning error determines the impact on the contrast, making it very difficult to compare it with other works. Ruane et al.~2022\mbox{
\cite{2022SPIE12180E..24R} }\hskip0pt
show the contour of $\pi/40$ radian phase errors for their manufactured mask near the central singularity. Llop-Sayson et al.~2024\mbox{
\cite{2024SPIE13092E..1YL} }\hskip0pt
show that the fast axis error converges to $\sim0.02$ radians far away from the central singularity for their best mask. This appears to be higher than our measurement of $0.012$ radians RMS, but the relative impact remains difficult to compare. We have simulated the impact of our patterning error on the contrast and discuss this in Section~\ref{sec:performance_sim}.
}

\DIFadd{The current patterning }\DIFaddend quality is primarily limited by the stability of the calibration, as the residuals remain clearly systematic. For example, we have confirmed that temperature and humidity variations in the clean room can influence this calibration. To mitigate this, a small calibration sample is manufactured immediately before writing the \DIFdelbegin \DIFdel{real }\DIFdelend \DIFaddbegin \DIFadd{final }\DIFaddend sample. Nevertheless, we still observe substantial fluctuations in the quality of the manufactured patterns. In the future, we aim to operate the direct-write machine in a temperature-stabilized environment. We also see that the fast axis error tends to increase near the central singularity, as shown in the right panel of Fig.~\ref{fig:1tr-writing-result}. However, this may also be the result of imperfect alignment of the \DIFaddbegin \DIFadd{microscope }\DIFaddend images with respect to each other.

\DIFdelbegin \DIFdel{To evaluate whether the achieved patterning quality meets the requirements for our application, we simulated the performance of the manufactured imperfect VVC using }\texttt{\DIFdel{hcipy}}
\DIFdel{~\mbox{
\cite{2018SPIE10703E..42P} }\hskip0pt
and the retrieved fast-axis error shown in Fig.~\ref{fig:1tr-writing-result}. The resulting coronagraph provides a contrast of $5 \times 10^{-8}$ at $3\,\lambda/D$ without wavefront control, and better than $10^{-9}$ after applying Electric Field Conjugation (EFC) in a simulated THD2 setup. Still, we hope to further improve the patterning quality in the near future. 
}

\DIFdelend 

The free parameters of the direct-write machine were also optimized for the manufacturing of polarization gratings. Similarly, a set of calibration samples was manufactured and we chose the settings with the lowest power in unwanted diffraction orders. \DIFdelbegin \DIFdel{We specifically minimized the presence of half-orders, which can limit the performance of the coronagraph. }\DIFdelend This will be discussed in more detail in \DIFdelbegin \DIFdel{section}\DIFdelend \DIFaddbegin \DIFadd{Section}\DIFaddend ~\ref{sec:ff_diff}.

 \subsection{Polarization leakage}

 Achieving broadband performance with minimal polarization leakage requires the coronagraphic mask to function as a half-wave retarder over a wide wavelength range. This can, for example, be achieved by stacking multiple layers of self-aligning twisted retarders \cite{2013OExpr..21..404K} or multiple rotated \DIFdelbegin \DIFdel{LCP masks \mbox{
\cite{Mawet2010SPIE_VVC_sensitivity}}\hskip0pt
}\DIFdelend \DIFaddbegin \DIFadd{untwisted LCP masks \mbox{
\cite{Mawet2010SPIE_VVC_sensitivity, 9843574}}\hskip0pt
}\DIFaddend . While the multiple \DIFdelbegin \DIFdel{uniform }\DIFdelend \DIFaddbegin \DIFadd{untwisted }\DIFaddend layer approach is simpler from a liquid-crystal processing perspective, it introduces more complex interface challenges and requires very accurate alignment of the components \DIFdelbegin \DIFdel{\mbox{
\cite{Mawet2011SPIE_vvc_review, 2019JOSAB..36D..13S}}\hskip0pt
}\DIFdelend \DIFaddbegin \DIFadd{\mbox{
\cite{Mawet2011SPIE_vvc_review, 2019JOSAB..36D..13S, 9843574}}\hskip0pt
}\DIFaddend . In this work, we use self-aligning twisted retarders to achieve broadband performance. This section reports on the development of a procedure for manufacturing VVCs with good broadband performance. 

The birefringence of the LCP material was first characterized by fabricating a prototype and measuring it using a J.A.~Woollam Co.\ RC2 spectroscopic ellipsometer, which retrieves the full Mueller matrix as a function of wavelength. The measured birefringence was subsequently used as input to a numerical optimization routine to determine the best-performing retarder design. Throughout this work, we employ designs consisting of three layers of self-aligning twisted retarders (3TR). The free parameters in the optimization are the twist angles and thicknesses of each retarder layer. The optimization is performed using a basin-hopping algorithm and explicitly accounts for expected manufacturing tolerances. To achieve uniform retardance across the full aperture, the spin-coater must be operated at high rotation speeds \DIFaddbegin \DIFadd{($>$1000 revolutions per minute)}\DIFaddend , resulting in thin LCP layers \DIFaddbegin \DIFadd{($\sim$400-700 nm) }\DIFaddend for each spin-coating step. We observed that the thickness of each successive spin-coated LCP sub-layer decreases as additional layers are deposited, likely due to changes in surface tension. This systematic decrease in thickness was therefore calibrated as a function of the layer number and incorporated into the design process.
\DIFdelbegin \DIFdel{After several design iterations, we obtained a broadband retardance profile that is not expected to limit the performance of multi-grating VVCs (mgVVCs).
}\DIFdelend 

\begin{figure}
\label{fig:leakage-wavelength} 
\centering
\DIFdelbeginFL 
\DIFdelendFL \DIFaddbeginFL \includegraphics[width = \linewidth]{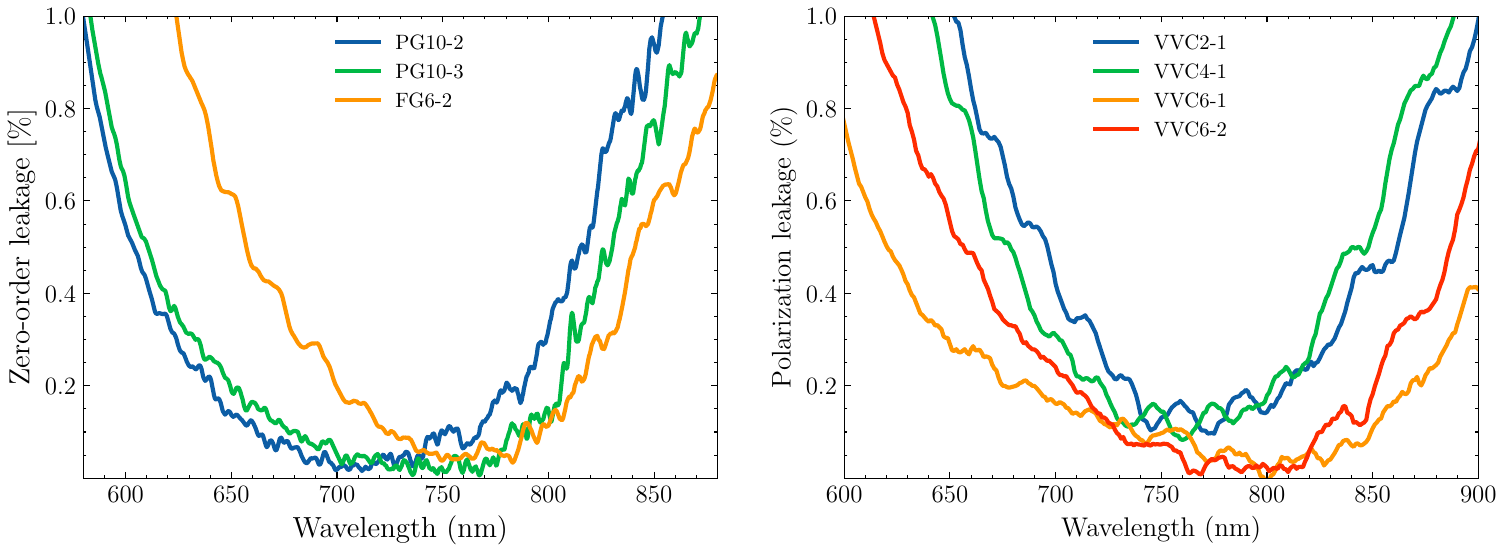}
\DIFaddendFL \caption[Leakages as function of wavelength]
{
\DIFdelbeginFL 
\DIFdelendFL Measured polarization leakage as a function of wavelength for different manufactured components. The left figure shows the measured zero-order leakage for polarization gratings while the right figure shows the measured polarization leakage for the manufactured vortex coronagraphs.}
\DIFaddbeginFL \label{fig:leakage-wavelength}
\DIFaddendFL \end{figure}

The polarization leakage of the fabricated optics was then measured. For the vortex coronagraphs, this measurement was performed using the spectro-polarimetric ellipsometer by retrieving the Mueller matrix of the transmitted light as a function of wavelength over the central 3~mm of the optic. For the polarization gratings, the leakage was measured in the zeroth diffraction order using a spectrometer. In this configuration, the polarization leakage is transmitted, while the diffracted orders are blocked and therefore not detected by the spectrograph. The measured zeroth-order signal thus includes both true polarization leakage and transmission arising from other imperfections in the optic, such as retardance non-uniformities, defects, and fast-axis errors. \DIFdelbegin \DIFdel{Results for }\DIFdelend \DIFaddbegin \DIFadd{For the 30 $\mu$m pitch forked grating (FG6-2) we had to insert circular polarizers in the beam to avoid the first diffraction order being measured for wavelengths $<$700 nm. The results for this component may still be slightly affected by this at these wavelengths. The results of }\DIFaddend the most recent batch of manufactured optics are shown in Fig.~\ref{fig:leakage-wavelength}.

The measured leakage curves are \DIFdelbegin \DIFdel{shifted slightly }\DIFdelend \DIFaddbegin \DIFadd{slightly shifted }\DIFaddend towards the red \DIFdelbegin \DIFdel{compared }\DIFdelend \DIFaddbegin \DIFadd{in comparison }\DIFaddend to the design predictions, indicating that the deposited layers are marginally thicker than anticipated during the spin-coating process. The best-performing vortex coronagraph has a polarization leakage of $<0.5\%$ over a bandwidth of approximately 40\% and $<0.2\%$ over a bandwidth of approximately 20\%. The best polarization grating shows an average polarization leakage of $3 \times 10^{-4}$ over a 10\% bandwidth and $8 \times 10^{-4}$ over a 20\% bandwidth. The small offset in central wavelength between the vortex and grating measurements may be the result of a wavelength calibration error in one of the two measurement instruments. While significant variability in leakage performance is observed between different manufacturing batches, good consistency is achieved among components fabricated within the same manufacturing cycle. The oscillatory features visible in the leakage curves are caused by interference fringes arising from the LCP to air interface.

\DIFaddbegin \subsection{\DIFadd{Performance simulation}}\label{sec:performance_sim}
\begin{figure}
    \centering
    \includegraphics[width=\linewidth]{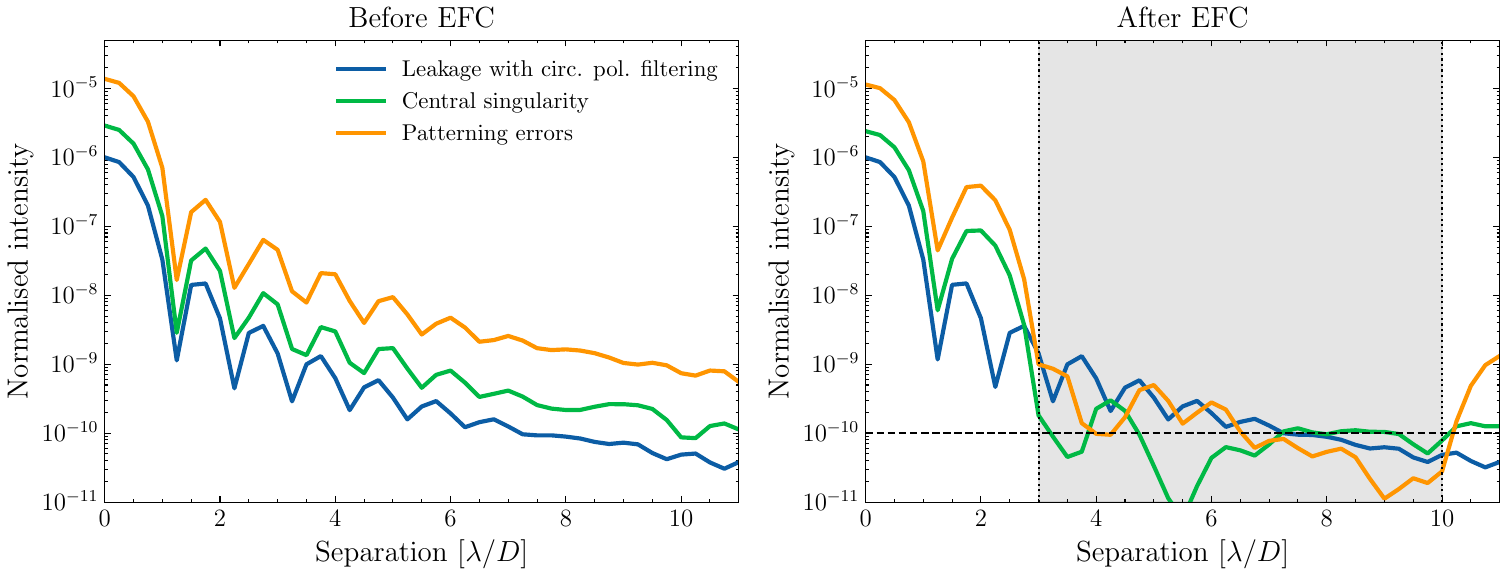}
    \caption{\DIFaddFL{Simulated impact on the contrast budget from various error terms for the manufactured charge 6 VVC at the THD2 for a 10\% bandwidth. The left figure shows the impact on the raw contrast, while the right figure shows the impact after dark hole digging between 3 and 10 $\lambda/D$ with Electric Field Conjugation (EFC).}}
    \label{fig:performanc_sim_vvc6}
\end{figure}

\DIFadd{To evaluate whether the manufactured components meet the requirements, we have performed simulations of the expected impact on the contrast at the THD2 bench both before and after wavefront control. These simulations were performed using hcipy\mbox{
\cite{2018SPIE10703E..42P} }\hskip0pt
and consisted of the following aspects: A circular input aperture of diameter ($D$) 8.23 mm was defined. Subsequently, we applied the shape of the first deformable mirror (DM) to the electric field, with a diameter of 9.9 mm and 34 actuators across. Next, we Fresnel propagate to the second DM, which is located at a distance of 26.9 mm\mbox{
\cite{2018SPIE10706E..2OB} }\hskip0pt
from the pupil and has the same shape and number of actuators as the first DM. We backpropagate to the pupil and use the hcipy multi-scale coronagraph implementation to simulate the imperfect VVC. This is then propagated through the Lyot stop with a diameter of 7.9 mm and we simulate the resulting focal plane image. All simulations are done for a 10\% bandwidth centered around a wavelength ($\lambda$) of 780 nm at five discrete wavelengths. For the wavefront control we assume that we have access to the noiseless electric field in the focal plane and use electric field conjugation (EFC) \mbox{
\cite{Give'on2007ClosedLoopDM} }\hskip0pt
to minimize the total intensity in the circular dark hole region between 3 and 10 $\lambda/D$. Fig.~\ref{fig:performanc_sim_vvc6} shows the simulated impact of the various imperfections on the contrast before and after wavefront control. The patterning error was modeled by adding the retrieved azimuthal fast axis error from Fig. \ref{fig:1tr-writing-result} to the focal plane phase mask. The central singularity was modeled as a phase error with a constant phase in a central region of $6 \mu$m in diameter instead of the desired vortex phase. All simulations assume circular polarization filtering with an average extinction of $10^{-3}$ across the bandpass. We see that patterning error is the biggest contributor to the contrast budget before wavefront control, with an estimated impact of $\sim 10^{-8}$ at $5\,\lambda/D$ without wavefront control, which goes down to a few times $10^{-10}$ after dark hole digging. However, correcting this takes up degrees of freedom on the DM and is not ideal. The contribution from the polarization leakage starts to dominate after wavefront control, as it is effectively incoherent and is not removed by EFC.
}

\DIFaddend \section{Assembly of multi-grating vector vortex coronagraphs}
\begin{figure}[htbp]
    \centering
    \DIFdelbeginFL 
\DIFdelendFL \DIFaddbeginFL \includegraphics[width=\linewidth]{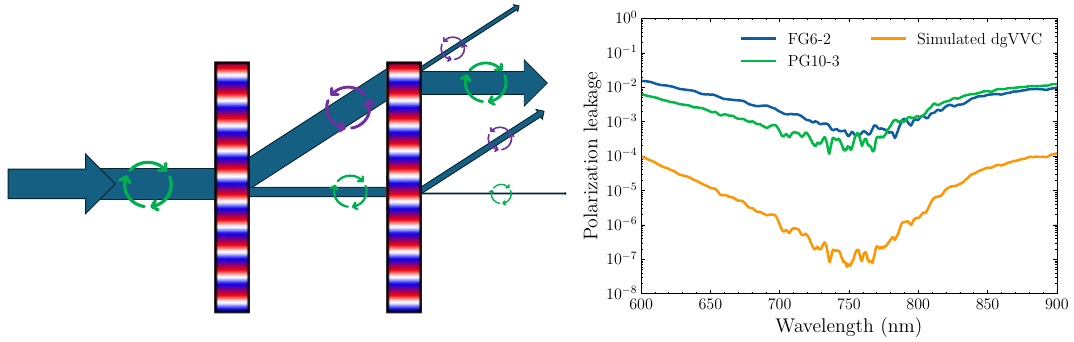}
    \DIFaddendFL \caption{\textbf{Left:} Illustration of the concept of the double-grating VVC, adapted from Doelman et al.\DIFaddbeginFL \DIFaddFL{~}\DIFaddendFL 2020\cite{2020PASP..132d5002D}. In reality, the two gratings can be glued together, resulting in no separation between the beams of both polarization states. \textbf{Right:} Polarization leakage as a function of wavelength for a manufactured forked grating and polarization grating, along with the first-order polarization leakage that the dgVVC should have when combining these elements.}
    \label{fig:dgVVC_concept}
\end{figure}

The leakage performance provided by the \DIFdelbegin \DIFdel{MTR }\DIFdelend \DIFaddbegin \DIFadd{3TR }\DIFaddend design is insufficient to reach the deep contrasts required for exo-Earth imaging without polarization filtering \DIFaddbegin \DIFadd{with a regular VVC}\DIFaddend . In this section we report on the results of developing mgVVCs that should be able to operate in broadband \DIFdelbegin \DIFdel{and without polarization filtering}\DIFdelend \DIFaddbegin \DIFadd{light and without being sandwiched between circular polarizers and therefore avoiding the loss of planet throughput}\DIFaddend . The concept of the mgVVC was \DIFdelbegin \DIFdel{demonstrated }\DIFdelend \DIFaddbegin \DIFadd{introduced }\DIFaddend in Doelman et al.~(2020)\cite{2020PASP..132d5002D}, and a sketch of the double-grating VVC (dgVVC) concept is illustrated in Fig.~\ref{fig:dgVVC_concept}. The idea is that the first grating, which is combined with a vortex into a forked grating, diffracts the main \DIFdelbegin \DIFdel{"vortexed" }\DIFdelend \DIFaddbegin \DIFadd{``vortexed'' }\DIFaddend beam off-axis, which is diffracted back on-axis by the second grating. On the other hand, the leakage from the first grating is diffracted off-axis by the second grating (and analogous for the leakage of the second grating), physically separating it from the main beam. While there is a distance between the grating in the illustration to more clearly visualize the concept, in practice they are glued together, which results in no separation between the polarization states of the main beam \cite{2020PASP..132d5002D}.

We have manufactured a dgVVC prototype at \DIFdelbegin \DIFdel{Colorlink }\DIFdelend \DIFaddbegin \DIFadd{ColorLink }\DIFaddend Japan. We used a grating pitch of 10 $\mu m$ for both the forked grating and the second polarization grating in the dgVVC. This is the minimum pitch that is currently achievable with good writing quality. The parameters of the direct-write machine were optimized for maximum diffraction efficiency and minimum power in non-desired diffraction orders. The dgVVC shown here used \DIFdelbegin \DIFdel{an older run }\DIFdelend \DIFaddbegin \DIFadd{the older set }\DIFaddend of manufactured components \DIFaddbegin \DIFadd{in August 2025}\DIFaddend , with worse leakage performance as the most recent batch shown in \DIFdelbegin \DIFdel{Fig.~\ref{fig:leakage-wavelength} }\DIFdelend \DIFaddbegin \DIFadd{Figures ~\ref{fig:leakage-wavelength} and ~\ref{fig:dgVVC_concept}}\DIFaddend .

  \subsection{Alignment \& assembly}
\begin{figure}
\centering
\DIFdelbeginFL 
\DIFdelendFL \DIFaddbeginFL \includegraphics[width = 1\linewidth]{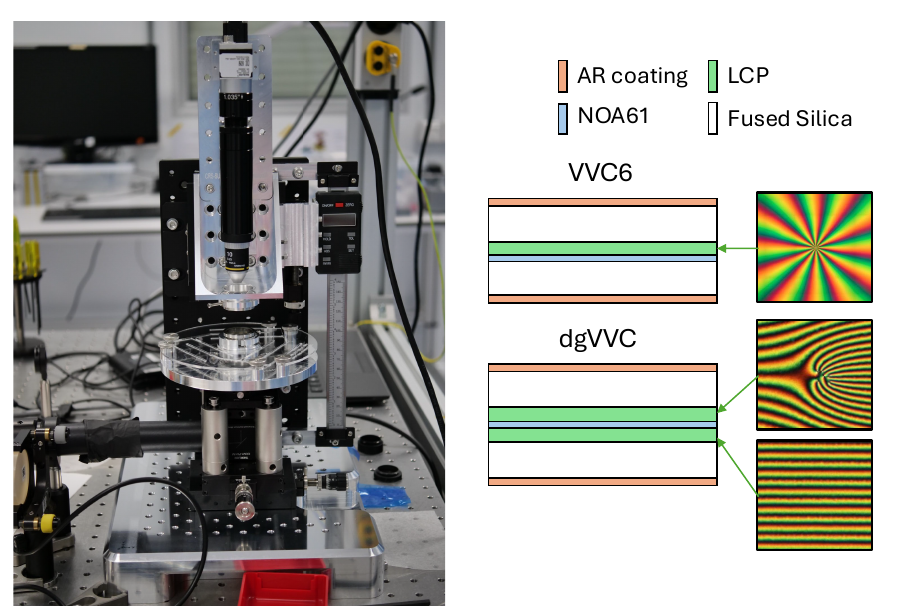}
\vspace{10pt}
\DIFaddendFL \caption[Jig]
{
\DIFdelbeginFL 
\DIFdelendFL \DIFaddbeginFL \textbf{\DIFaddFL{Left:}} \DIFaddendFL Alignment and bonding jig for multi-component liquid-crystal mask assemblies, developed and validated at cosine \DIFdelbeginFL \DIFdelFL{.
}\DIFdelendFL \DIFaddbeginFL \DIFaddFL{Remote Sensing. }\textbf{\DIFaddFL{Right:}} \DIFaddFL{Schematic overview of the liquid-crystal mask assemblies (LCMAs).
}\DIFaddendFL }
\DIFaddbeginFL \label{fig:Jig}
\DIFaddendFL \end{figure}

\begin{figure}
\centering
\DIFdelbeginFL 
\DIFdelendFL \DIFaddbeginFL \includegraphics[width = \textwidth]{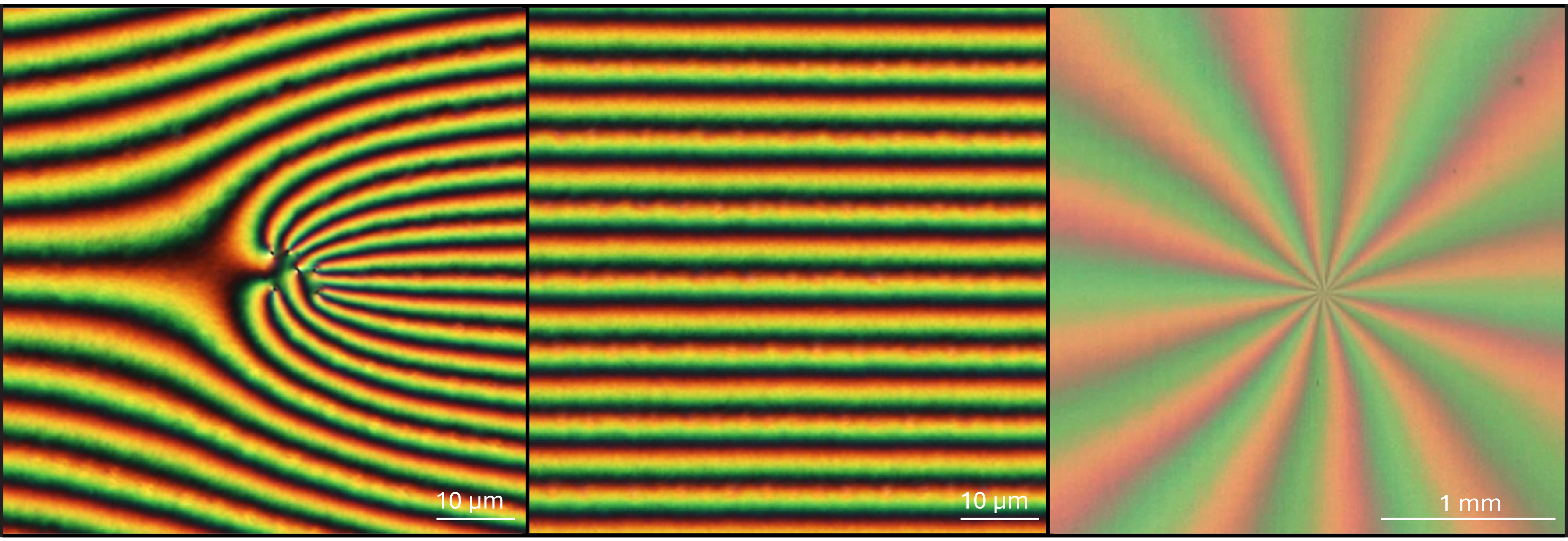}
\DIFaddendFL \caption[Gluing] 
{\label{fig:gluing} 
\textbf{Left:} Microscope image between \DIFdelbeginFL \DIFdelFL{crossed-polarizers }\DIFdelendFL \DIFaddbeginFL \DIFaddFL{crossed polarizers }\DIFaddendFL of a manufactured charge 6 forked grating \DIFaddbeginFL \DIFaddFL{(FG10-1)}\DIFaddendFL . \textbf{Middle:} Microscope image between \DIFdelbeginFL \DIFdelFL{crossed-polarizers }\DIFdelendFL \DIFaddbeginFL \DIFaddFL{crossed polarizers }\DIFaddendFL of a manufactured polarization grating \DIFaddbeginFL \DIFaddFL{(PG10-1)}\DIFaddendFL . \textbf{Right:} Microscope image between \DIFdelbeginFL \DIFdelFL{crossed-polarizers }\DIFdelendFL \DIFaddbeginFL \DIFaddFL{crossed polarizers }\DIFaddendFL of the assembled double-grating VVC, consisting of the charge 6 forked grating and the polarization grating shown in the left and middle panel. This image was taken using a different microscope than the two images on the left.
}
\end{figure}

  The components were then assembled at cosine \DIFaddbegin \DIFadd{Remote Sensing B.V. }\DIFaddend in a clean room environment. An alignment jig was developed and validated\DIFdelbegin \DIFdel{at cosine, seen in Fig.~\ref{fig:Jig}, }\DIFdelend \DIFaddbegin \DIFadd{, }\DIFaddend which can accurately align and bond multiple liquid-crystal masks into liquid-crystal mask assemblies (LCMAs), while also \DIFaddbegin \DIFadd{being }\DIFaddend capable of aligning a chromium mask into the stack. \DIFdelbegin \DIFdel{The jig }\DIFdelend \DIFaddbegin \DIFadd{An image of the jig can be seen in Fig.~\ref{fig:Jig}. It }\DIFaddend was designed to achieve a clocking accuracy of $<1$ degrees and an XY placement of  $\pm1$ $\mu m$. After alignment the components can be glued together with NOA61 glue, which is a UV-curing optical adhesive. This custom alignment jig also consists of an integrated crossed-polarizers microscope. This function allows \DIFaddbegin \DIFadd{us }\DIFaddend to capture polarization images of both the individual components, as well as the assembled LCMA\DIFdelbegin \DIFdel{, as a reference a }\DIFdelend \DIFaddbegin \DIFadd{. An image of the assembled }\DIFaddend dgVVC is shown in Fig.~\DIFdelbegin \DIFdel{\ref{fig:microscope-image}. }\DIFdelend \DIFaddbegin \DIFadd{\ref{fig:gluing}. For this component, the gratings cancel each other, and it looks very similar to a regular VVC. }\DIFaddend The slight curve in the vortex pattern is due to imperfect clocking of the gratings with respect to each other. We have simulated the observed image and find that it can be explained by a clocking error of about 0.05 degrees.

 \subsection{Far-field diffraction tests}\label{sec:ff_diff}
\begin{figure}
    \centering
    \includegraphics[width=0.7\linewidth]{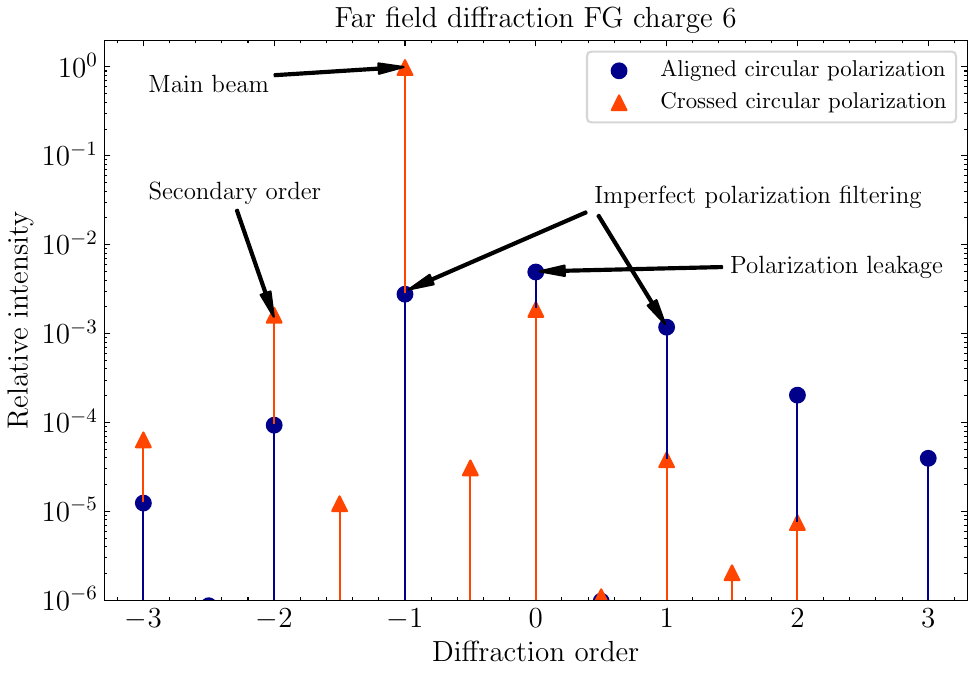}
    \caption{Far-field diffraction measurements of a charge 6 forked grating (\DIFdelbeginFL \DIFdelFL{FG}\DIFdelendFL \DIFaddbeginFL \DIFaddFL{FG6-1}\DIFaddendFL ), showing the relative intensity as a function of diffraction order. The input beam was circularly polarized and the intensity is shown both for and aligned and crossed circular polarization state of the analyzer. Explanations of the origin of various terms are annotated in the figure.}
    \label{fig:ff-measurements-fg}
\end{figure}

A perfect polarization grating with half-wave retardance will diffract all the light in the $\pm$1 diffraction orders, split according to circular polarization. Retardance deviations from half-wave will result in leakage in the zeroth order. Additionally, fast axis errors may result in higher order and half order diffracted terms that can fold back into the Lyot stop and cause additional stellar leakage. To verify this, a setup has been constructed to measure the far-field diffraction patterns of the (forked) gratings and assembled dgVVC. This setup operates at a wavelength of 780 nm and uses a polarizer and quarter-wave plate to circularly polarize the input beam. We also have a circular polarizer as analyzer which can be rotated to measure both left- and right-handed circular polarization. For all components we observe copies of the pupil at multiples of the diffraction order. The \DIFdelbegin \DIFdel{results }\DIFdelend \DIFaddbegin \DIFadd{total intensity in each diffraction order }\DIFaddend for the charge 6 forked grating used in the dgVVC is shown in Fig.~\ref{fig:ff-measurements-fg}. \DIFaddbegin \DIFadd{Note that the beam was not focused on the fork and thus does not get the vortex pattern imprinted on it.
}\DIFaddend 

\begin{figure}
    \centering
    \includegraphics[width=0.7\linewidth]{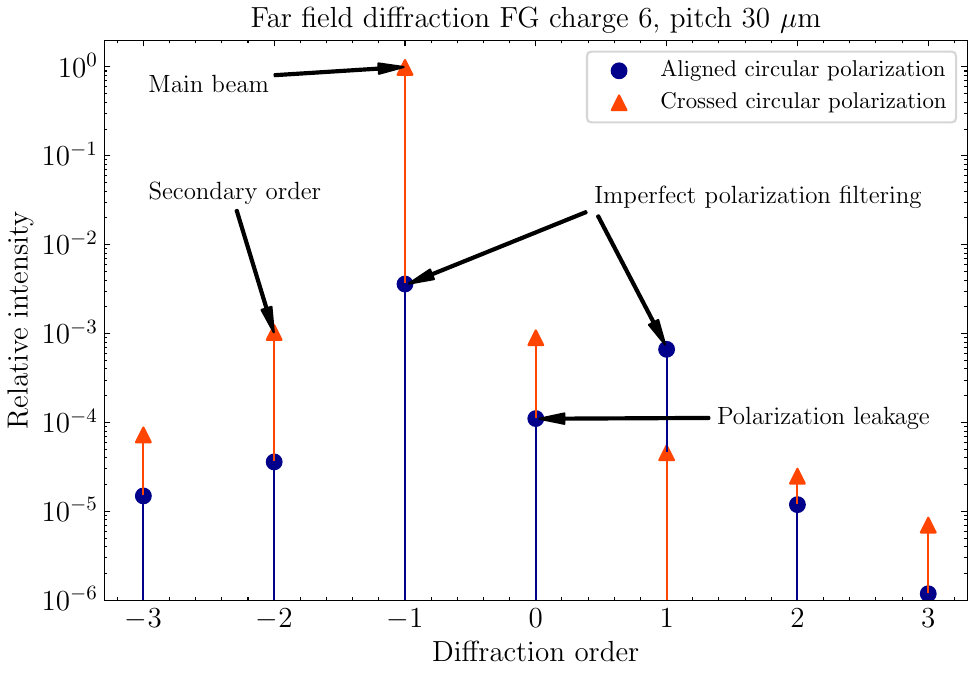}
    \caption{Same as Fig.~\ref{fig:ff-measurements-fg}, but now for one of the most recently manufactured forked gratings \DIFaddbeginFL \DIFaddFL{with a 30 $\mu$m pitch (FG6-2)}\DIFaddendFL .}
    \label{fig:ff-measurements-fg-d3}
\end{figure}

These measurements show a polarization leakage, which is given by the 0th diffraction order with aligned circular polarization, of $\sim5 \times 10^{-3}$ for this component. This is significantly lower for the \DIFdelbegin \DIFdel{newly }\DIFdelend \DIFaddbegin \DIFadd{newer batch of }\DIFaddend manufactured components, for which the far-field diffraction measurement is shown in Fig.~\ref{fig:ff-measurements-fg-d3}, confirming the low leakage shown in Fig.~\ref{fig:leakage-wavelength}. The measurements also show that the alignment of the circular polarizers or retardance of the quarter-waveplate is not perfect. First, we see that we do not create one circular polarization state, shown by the presence of the +1 order in aligned circular polarization. Additionally, the analyzer reaches a filtering of $\sim3\times 10^{-3}$, shown by the presence of the aligned circular polarization term at diffraction order -1.

\begin{figure}
    \centering
    \includegraphics[width=0.7\linewidth]{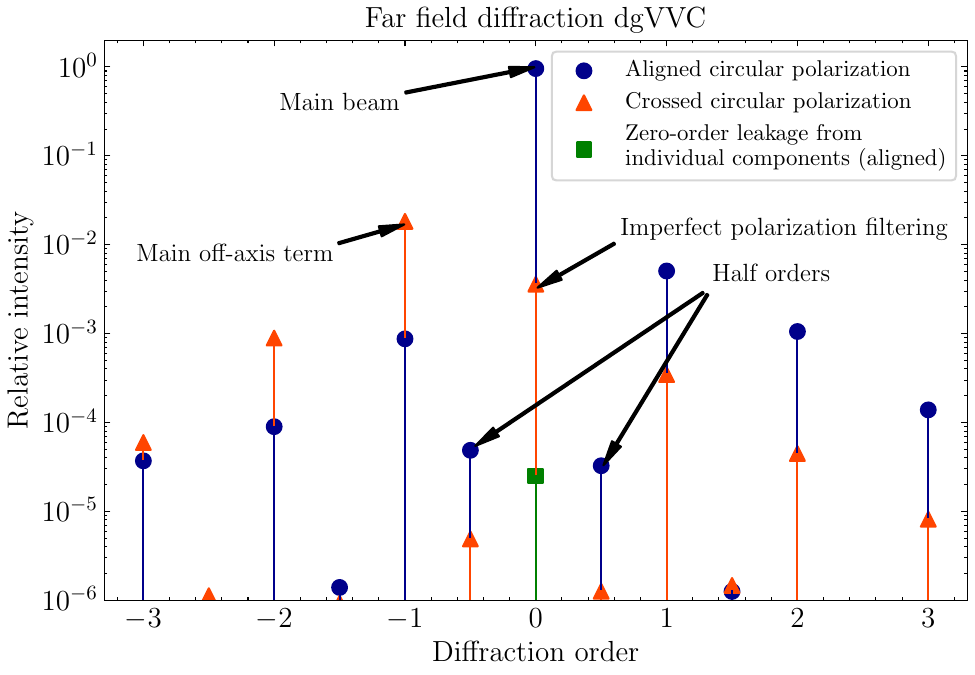}
    \caption{Far-field diffraction measurements of the \DIFaddbeginFL \DIFaddFL{assembled }\DIFaddendFL double-grating VVC, showing the relative intensity as a function of diffraction order. The input beam was circularly polarized and the intensity is shown both for and aligned and crossed circular polarization state of the analyzer \DIFaddbeginFL \DIFaddFL{with respect to the input circular polarization state}\DIFaddendFL . We have also plotted the zero-order leakage contribution to the main beam in aligned circular polarization, which we have simulated using the measured leakage for the individual components.}
    \label{fig:ff-measurements}
\end{figure}

\begin{figure}
    \centering
    \includegraphics[width=0.7\linewidth]{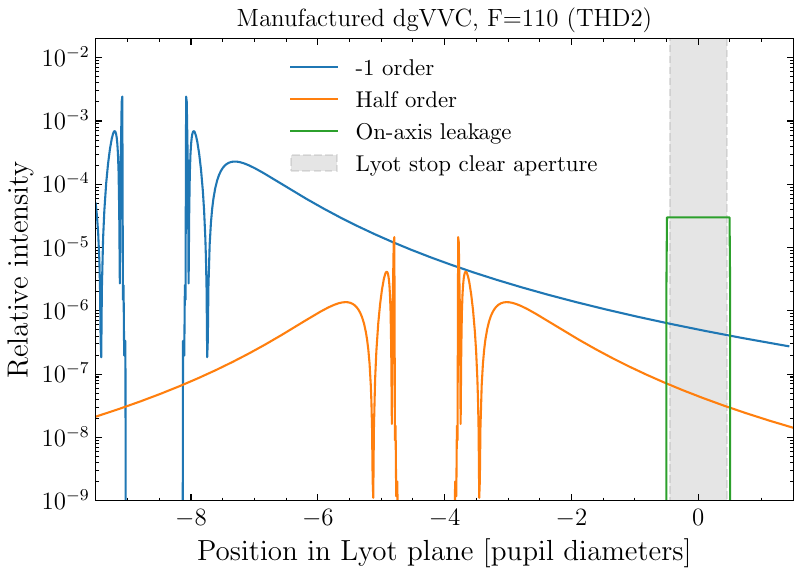}

    \caption{\textbf{\DIFaddFL{Left:}} \DIFaddFL{Simulated Lyot-plane intensity for the manufactured dgVVC and a circularly polarized input beam for a wavelength of 780 nm. We do not show terms that do not majorly contribute to the intensity in the Lyot stop clear aperture, such as the -1 order without the vortex phase imprinted.}}
    \label{fig:lyot_plane_intensity}
\end{figure}

We have also measured the far-field diffraction of the assembled dgVVC, which is shown in Fig.~\ref{fig:ff-measurements}. This shows that the main beam is indeed diffracted back on-axis. The second brightest feature is at the -1 order with opposite polarization state as the main beam. This is mainly the combination of 1) the beam diffracted by the forked grating and leaked by the second grating, and 2) the beam leaked by the forked grating and diffracted by the second grating. The second is no issue, as it is a regular \DIFdelbegin \DIFdel{pupil }\DIFdelend \DIFaddbegin \DIFadd{reimaged pupil that originated as zero-order leaked light in the forked grating }\DIFaddend and will be blocked by the Lyot stop. However, the first \DIFaddbegin \DIFadd{term }\DIFaddend has the vortex \DIFaddbegin \DIFadd{phase }\DIFaddend pattern imprinted on it, resulting in a ``ring of fire'' in the pupil plane that has wings that can leak into the clear aperture of the Lyot stop\DIFdelbegin \DIFdel{. Given the grating pitch of 10 $\mu m$, a spatial resolution $F\lambda$ of 86 $\mu$m at a wavelength of 783 nm, and assuming a conservative polarization leakage of 0.5\%, this results in an additional on-axis leakage term with an intensity of $3\times 10^{-7}$, which decreases to $\sim 10^{-9} $ at 3 $\lambda/D$.While this may not be an issue for the tests at the THD2 bench because of its F/110 beam, it should be considered for instruments with faster beams in the coronagraphic focal plane.
Decreasing the grating pitch or improving the fast axis writing quality can mitigate this. However, this main off-axis term has the opposite polarization state as the main beam and can thus be removed using polarization filtering. Additionally, the significantly reduced leakage for newer components will help mitigate this issue.
}\DIFdelend \DIFaddbegin \DIFadd{, as shown in Fig.~\ref{fig:lyot_plane_intensity}.
}\DIFaddend 

Initial components also showed strong half-orders, but this issue has been mitigated by optimizing the settings and writing direction of the direct-write machine. These half-orders can be detrimental to the contrast as they are physically closer to the Lyot stop. This leakage term can also not be removed with \DIFdelbegin \DIFdel{polarization filtering}\DIFdelend \DIFaddbegin \DIFadd{a circular analyzer}\DIFaddend , as it has the same polarization state as the main beam. \DIFdelbegin \DIFdel{Fig.~\ref{fig:ff-measurements} shows that these half-orders have intensities of $\sim5\times 10^{-5}$ for the manufactured dgVVC. This results in additional on-axis leakage of $5 \times \sim 10^{-8}$, which goes down to well below $10^{-9}$ at 3$\lambda/D$. Newly manufactured components do not show these half orders }\DIFdelend \DIFaddbegin \DIFadd{The new batch of manufactured components does not have detectable half orders with intensities of $>10^{-6}$}\DIFaddend , as shown in Fig.~\ref{fig:ff-measurements-fg-d3}. We do not observe any other significant scattered light between the diffraction orders originating from imperfect gratings. \DIFdelbegin \DIFdel{The far-field diffraction measurementsagree well with our simulations and the grating quality should not }\DIFdelend \DIFaddbegin \DIFadd{Using these measurements, we have simulated the Lyot-plane intensities for the manufactured dgVVC and the results are shown in Fig.~\ref{fig:lyot_plane_intensity}. 
}

\subsection{\DIFadd{Performance simulation}}
 \DIFadd{Here, we simulate the impact of the grating quality and the assembly on the achievable contrast at the THD2 bench. Given the grating pitch of 10 $\mu m$, an F/110 beam and a wavelength of 783 nm, the simulated Lyot plane intensities are shown in Fig. \ref{fig:lyot_plane_intensity}. This shows that the on-axis leakage, which can not be filtered for the dgVVC, will be the main form of leakage for the manufactured component at the THD2. We see that the -1 diffraction order results in an additional on-axis leakage term with an intensity of $5\times 10^{-7}$, which decreases to $\sim 2 \times 10^{-9} $ at 3 $\lambda/D$. This main off-axis term has the opposite polarization state as the main beam and could thus be removed using a circular analyzer. Both the on-axis leakage as well as the -1 order would be reduced for a dgVVC assembled using the newest components. Still, high-order diffraction terms in the same polarization state as the main beam are seen at an intensity of $\sim10^{-3}$ as a result of patterning imperfections, as seen in Fig. \ref{fig:ff-measurements-fg-d3}. Combined with the significantly improved leakage performance of newer components, the contribution from the on-axis leakage and this off-axis term would be similar. For instruments with a faster beam near the coronagraphic focal plane or targeting shorter wavelengths, the off-axis terms start to dominate over the on-axis leakage. We conclude that while the grating patterning quality will not limit our tests of the dgVVC prototype at the THD2, they may }\DIFaddend be a limiting factor for \DIFdelbegin \DIFdel{the performance of the manufactured dgVVC .
}\DIFdelend \DIFaddbegin \DIFadd{an HWO-like instrument with a smaller F-number at the coronagraphic focal plane. Improving the grating patterning quality or decreasing the grating pitch could help mitigate the impact of these.
}\DIFaddend 

 \DIFaddbegin \DIFadd{We have also simulated the impact of the slight rotational misalignment between the forked grating and the polarization grating, which results in an imperfect vortex phase pattern. We find that this rotational misalignment leads to a small shift in the location of the pupil in the Lyot plane. This shift is flipped for both circular polarization states and scales linearly with wavelength. For our misalignment error and the Lyot stop size at THD2, there is no additional light leaked through the Lyot stop. We find a tolerance to the clocking error of $\sim0.1$ degrees before light starts to leak inside the Lyot stop. 
 }

 \DIFadd{We also find in our simulations that there is a reduced impact on the contrast from the central singularity of the dgVVC compared to the regular VVC by about a factor of three. This is the result of the leaked light from the singularity being diffracted away by the second grating. The dgVVC may thus, depending on the wavelength and F-number of the instrument, not need a central opaque dot.
}

\DIFaddend \section{High-contrast results}
 The experimental validation of the manufactured VVCs and dgVVCs was performed at the \DIFdelbegin \DIFdel{Tr\`es Haute Dynamique 2 (}\DIFdelend THD2 \DIFdelbegin \DIFdel{) }\DIFdelend bench at LIRA, Observatoire de Paris. The THD2 bench employs a multi–deformable mirror (DM) configuration to correct both phase and amplitude aberrations\cite{2018SPIE10706E..2OB}. It has been used to evaluate and compare wavefront sensing and control (WFS\&C) techniques under equivalent experimental conditions, achieving raw contrast levels on the order of $10^{-8}$ to $10^{-9}$, despite operation in air. In this study, the \DIFdelbegin \DIFdel{WF}\DIFdelend \DIFaddbegin \DIFadd{WFS}\DIFaddend \&C strategy combines pairwise wavefront sensing \cite{Give'on2011PairwiseDeformableMirror} with \DIFdelbegin \DIFdel{Electric Field Conjugation (EFC ) control\mbox{
\cite{Give'on2007ClosedLoopDM}}\hskip0pt
}\DIFdelend \DIFaddbegin \DIFadd{EFC control}\DIFaddend . 
 Pairwise sensing is used to estimate the complex electric field in the focal plane, while EFC computes deformable-mirror commands that iteratively minimize the stellar intensity within a specified region of the focal plane. This approach has demonstrated robust performance across multiple coronagraph architectures on the testbed \cite{Potier2020ComparingFocalPlane, Laginja2025ROMAN_WFS}.
 In addition, the THD2 has been used to characterize the performance of several coronagraphs, spanning highly chromatic designs, such as the four-quadrant phase-mask (FQPM\cite{Potier2020ComparingFocalPlane}) coronagraph, to more broadband concepts, including the dual-zone phase-mask (DZPM \cite{Delorme2016LaboratoryValidation}), six-level phase-mask coronagraph \cite{Patru2018InlabTestingSixlevel}, and wrapped vortex coronagraph (WVC \cite{Galicher2020AFamilyofPhaseMasks}).\\
 As part of the SUPPPPRESS project, the THD2 was upgraded with an adjustable broadband supercontinuum white-light fiber laser source (SuperK Fianium, NKT Photonics). The central wavelength and spectral bandwidth are set using a dedicated tunable filter, allowing bandwidths from 10 to 100 nm over a wavelength range of 500 to 850 nm on the bench.

The testbed allows several options for polarization filtering. One approach consists of inserting a linear polarizer upstream of the camera to mitigate differential polarization effects, which have been measured to be at the level of a few $10^{-8}$ on the THD2 bench\cite{Baudoz2024PolarizationEffects}. Another approach, used for vectorial optical elements, relies on circular polarization filtering, implemented by placing a circular polarizer upstream of the coronagraph and a circular analyzer downstream.
These circular polarization components are composed of a linear polarizer combined with a custom achromatized quarter-wave plate. They are located in converging and diverging beams; however, because the numerical aperture of the bench is slow (F \DIFdelbegin \DIFdel{/D }\DIFdelend $\sim$ 110) at the coronagraph focus, this configuration is sufficient to achieve polarization filtering factors of $10^{4}$ – $10^{5}$. Despite the presence of anti-reflection coatings, these components appear to introduce ghost features in the focal plane that are not observed when inserting the linear polarizer in front of the camera or without any polarization filtering.

\begin{figure}[htbp]
     \centering
     \includegraphics[width=0.6\linewidth]{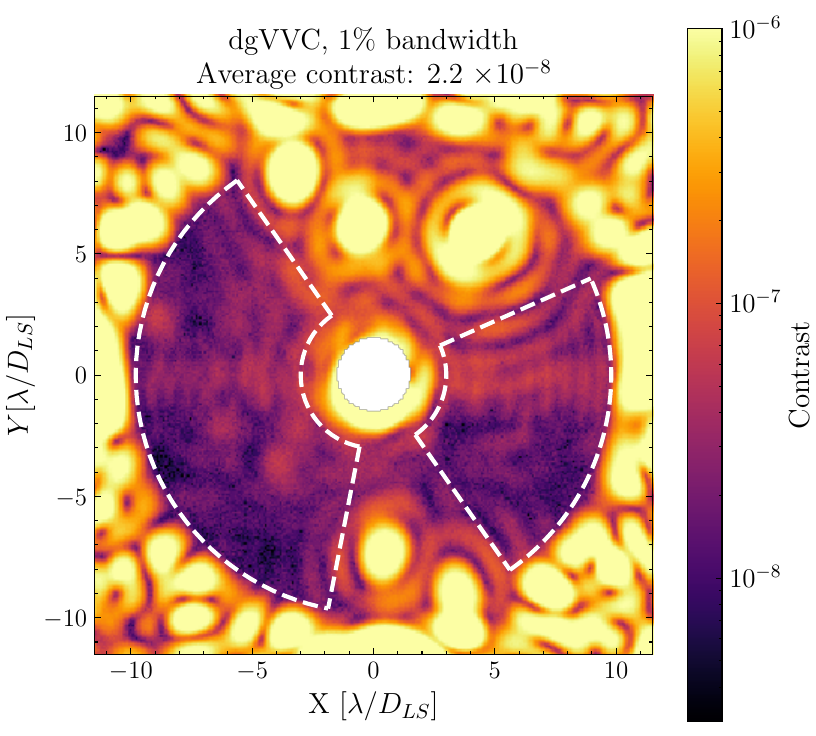}
     \caption{Focal plane image after dark hole digging for the assembled charge 6 dgVVC. We observe numerous ghost in the image which are not the result of the coronagraphic mask.}
     \label{fig:contrast_full}
 \end{figure}

\begin{figure}[htbp]
\centering
\includegraphics[width = \textwidth]{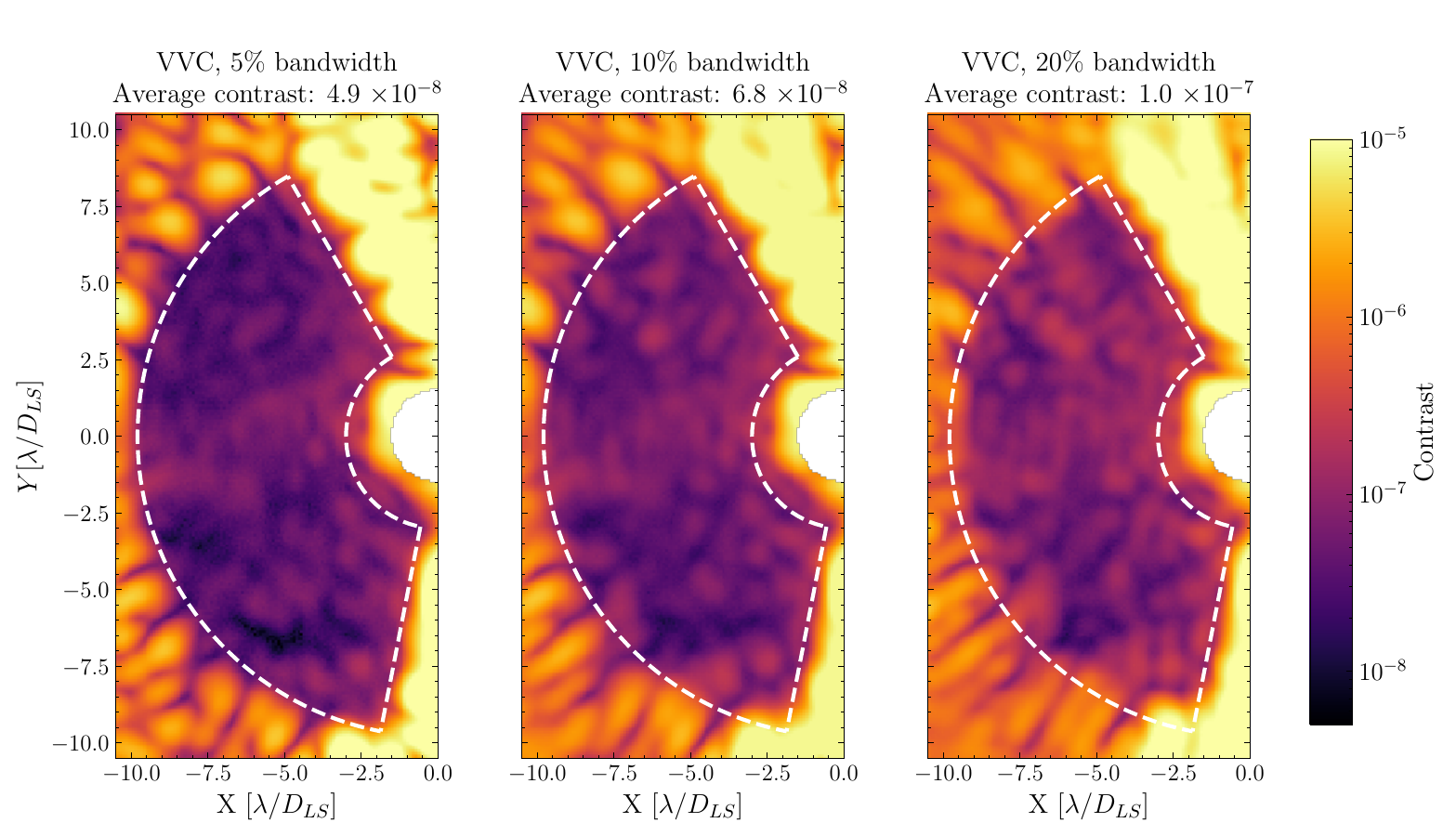}
\caption[Contrast curves] 
{\label{fig:vvc_contrast_bandwidth} 
Focal plane images after dark hole digging for the manufactured charge 6 VVC prototype \DIFaddbeginFL \DIFaddFL{(VVC6-1) }\DIFaddendFL as a function of bandwidth, centered around 780 nm. The encircled white area indicates the dark hole digging and scoring region.
}
\end{figure}

\begin{figure}[htbp]
\centering
\includegraphics[width = \textwidth]{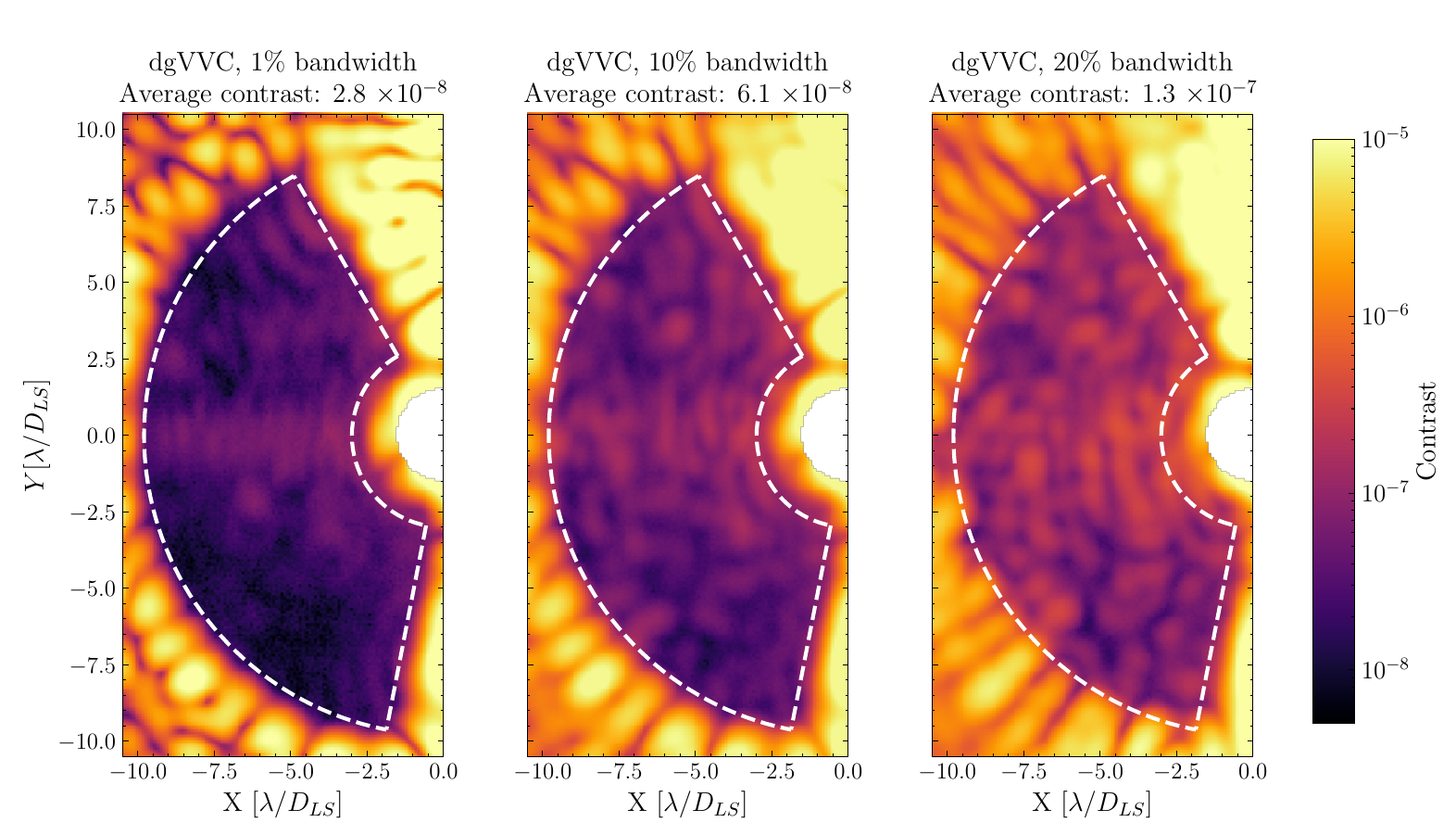}
\caption[Contrast curves] 
{\label{fig:dgvvc_contrast_bandwidth} 
Focal plane images after dark hole digging for the manufactured dgVVC prototype as a function of bandwidth, centered around 780 nm. The encircled white area indicates the dark hole digging and scoring region.
}
\end{figure}

We initially \DIFdelbegin \DIFdel{find }\DIFdelend \DIFaddbegin \DIFadd{used a full circular dark hole between 0 and 10 $\lambda/D$. However, we found }\DIFaddend that the dark hole digging is unstable for the regular VVC when using an inner working angle of 0 $\lambda/D$. We confirm that this behaviour can be reproduced in simulations and is the result of \DIFdelbegin \DIFdel{fast axis }\DIFdelend \DIFaddbegin \DIFadd{patterning }\DIFaddend errors in the manufactured component. We subsequently increase the inner working angle for dark hole digging to 3 $\lambda/D$. \DIFdelbegin \DIFdel{We qualitatively find that the dgVVC is more robust than the regular VVC, as it is for example less sensitive to the choice of the regularization parameter.  }\DIFdelend Fig.~\ref{fig:contrast_full} shows the resulting focal-plane image of the dgVVC after dark hole digging, along with the average contrast. This is measured by taking the average intensity in the scoring region (3-10 $\lambda/D$) and dividing by the peak intensity of an off-axis source at $\sim 7$ $\lambda/D$. 
 The ghost features visible in this image are associated with the \DIFdelbegin \DIFdel{use of polarization filtering}\DIFdelend \DIFaddbegin \DIFadd{quarter-wave plates}\DIFaddend . We have validated that these are indeed ghosts and not the result of the coronagraph, as they are at the exact same location when using a different coronagraph, rotating the mask or changing the wavelength. We exclude these ghosts in the calculation of the average contrast and the scoring region is shown with white dashed lines in Fig.~\ref{fig:contrast_full}. The measured \DIFdelbegin \DIFdel{monochromatic contrast }\DIFdelend \DIFaddbegin \DIFadd{contrast in a 1\% bandwidth }\DIFaddend of $2.2 \times 10^{-8}$ is close to the incoherent background limit of the testbed. This estimated contrast may still be slightly affected by Airy rings originating from ghost reflections. For the remainder of the tests, we therefore restrict the analysis to half dark holes located in the left scoring region. We also investigate the performance as a function of increasing bandwidth. The results for the charge-6 VVC and the dgVVC are shown in Fig.~\ref{fig:vvc_contrast_bandwidth} and Fig.~\ref{fig:dgvvc_contrast_bandwidth}, respectively. While it is difficult to determine what is the main limitation on the performance, we have identified the following sources as potential limitations:

\begin{itemize}
    
\item
\DIFdel{Fast-axis }\DIFdelend \DIFaddbegin \DIFadd{Patterning }\DIFaddend errors in the manufactured components\DIFadd{, which the simulations from Section~\ref{sec:performance_sim} showed to be the largest contributor for the VVC6-1 before EFC.}
    \DIFdelbegin \DIFdel{Since the Jacobian used in the EFC loop assumes an ideal vortex pattern, these non-idealities may lead to a sub-optimal correction.
    }\DIFdelend \item Polarization leakage. The dgVVC prototype was assembled using an earlier batch of components with higher polarization leakage than more recent samples, with an estimated on-axis leakage of $\sim 3 \times 10^{-5}$ in the same polarization state as the main beam. This leakage is expected to be reduced by \DIFdelbegin \DIFdel{at least }\DIFdelend \DIFaddbegin \DIFadd{about }\DIFaddend a factor of 100 for a dgVVC assembled with the newest components, as shown in Fig.~\ref{fig:dgVVC_concept}.
    \item \DIFaddbegin \DIFadd{Ghosts. The airy rings originating from the ghosts from components external to the coronagraph may impact the EFC solution.
    }\item \DIFaddend Chromaticity of the EFC solution. The EFC optimization was performed at the central wavelength, which may result in sub-optimal performance over larger bandwidths. Broadband estimators may help improve the broadband performance \cite{2016JATIS...2a1009G, 2021SPIE11823E..1QR}.
\end{itemize}

One of the main advantages of the dgVVC is its significantly reduced polarization leakage and its potential to be operated without \DIFdelbegin \DIFdel{polarization filtering}\DIFdelend \DIFaddbegin \DIFadd{circular polarizers}\DIFaddend , which would increase planet throughput by \DIFaddbegin \DIFadd{at least }\DIFaddend a factor of two. However, initial tests performed without \DIFdelbegin \DIFdel{circular polarization filtering of }\DIFdelend \DIFaddbegin \DIFadd{circularly polarizing }\DIFaddend the input beam were unsuccessful. In this configuration, the dgVVC functions as a charge $+6$ and charge $-6$ vortex for the two circular polarization states. Since the polarization states are not separated, the resulting focal-plane image is a combination of both vortex charges. In this regime, classical EFC using a single Jacobian does not provide a valid correction for both polarization states simultaneously \cite{2021SPIE11823E..1TM}. We therefore attempted to apply implicit EFC (iEFC) \cite{2023A&A...673A..28H} to the \DIFdelbegin \DIFdel{unfiltered dgVVC configuration. }\DIFdelend \DIFaddbegin \DIFadd{dgVVC without a circular polarizer. This algorithm uses an empirically calibrated Jacobian and can in principle work for both polarization states for the VVC simultaneously, albeit with a reduction in achieved contrast.\mbox{
\cite{2023A&A...673A..28H} }\hskip0pt
Additionally, the empirical calibration will account for any mask imperfections in the obtained Jacobian. }\DIFaddend However, these tests did not achieve deep contrast levels. This could be the result of the iEFC calibration or implementation, which was not optimized or validated \DIFdelbegin \DIFdel{without polarization filtering}\DIFdelend \DIFaddbegin \DIFadd{with a circular polarizer}\DIFaddend . Alternatively, residual imperfections in the coronagraph mask \DIFaddbegin \DIFadd{or interaction with surface errors in the optical system }\DIFaddend may lead to polarization-dependent aberrations, resulting in different optimal EFC solutions for the two polarization states. Polarization aberrations and effects are also known to be present on the THD2 bench itself \cite{Baudoz2024PolarizationEffects}. Future work will further investigate the performance of the components in dual-polarization mode, also taking into account polarization aberrations and exploring alternative dark-hole digging strategies. This may include approaches that separate the two circular polarization point spread functions. We note that even when mgVVCs are operated with \DIFdelbegin \DIFdel{polarization control}\DIFdelend \DIFaddbegin \DIFadd{a circular polarizer}\DIFaddend , the multi-grating concept can still \DIFdelbegin \DIFdel{provide a significant advantage by increasing }\DIFdelend \DIFaddbegin \DIFadd{increase }\DIFaddend the usable spectral bandwidth of the coronagraph.

\section{Environmental tests}
A separate part of the SUPPPPRESS project was to perform a basic environmental test campaign on representative liquid-crystal optics in order to demonstrate the suitability of these components for application in space. We have focussed on two aspects for space qualification. Firstly, we have assessed the tolerance of the optical components to $\gamma$- and UV-radiation, in particular with respect to the liquid-crystal layers. Secondly, we have designed and realized a preliminary mounting concept for the optical components and subjected the assembly to flight representative levels of vibration and thermal cycling in order to demonstrate the structural integrity of the assembly as well as the optical component itself. \DIFdelbegin \DIFdel{Tab.}\DIFdelend \DIFaddbegin \DIFadd{Table}\DIFaddend ~\ref{tab:environmental} summarizes the environmental tests that were performed together with their levels and duration and also lists the items under test and their specification.
As part of the $\gamma$- and UV-radiation tests we have performed pre- and post-test transmission measurements through the center of each optic with a beam diameter of $\approx$ 6 mm. In addition, we have taken pre- and post-test (polarized) microscope images of the liquid-crystal pattern. Similarly, we have taken microscope images of the bonding pads of the mounted optic prior and after the vibration and thermal cycling tests in order to check for damage or delamination at this critical interface. 

\begin{table}
    \centering
    \footnotesize
    \begin{tabular}{|p{1.6cm}|p{2.7cm}|p{2.4cm}|p{1.5cm}|p{1.7cm}|p{3.9cm}|}
         \hline
         \textbf{Test} & \textbf{Level} & \textbf{Duration} & \textbf{Total} & \textbf{Items} & \textbf{Item specification} \\
         \hline \hline
         $\gamma$-radiation & 791 rad/hour & 42.29 hours & 33.45 krad & 1-5 & \multirow{4}{3.9cm}{\scriptsize S1: 1AR-FS \par S2: 1AR-FS + 1TR test pattern \par S3: 1AR-FS + 1TR test pattern \par S4: 1AR-FS + 1TR forked grating \par S5: 1AR-FS-1TR double vortex assembly \par S6: 1AR-FS + 1TR single vortex capped with a FS-1AR \par S7: 1AR-FS + 3TR single vortex capped with a FS-1AR \par S8: 3-component grating assembly \par S9: FS-FS NOA61 glue test assembly} \\
         \cline{1-5}
         UV-radiation & equivalent to solar radiation at 1 AU & 20-45 hours &  & S4-S7, S9 & \\
         \cline{1-5}
         Vibration & Sine: 2-120 Hz @ \par \quad 16g in-plane \par \quad 22g out-of-plane \par Random & Sine: 2.3 min.\par Random: 2 min. &  & S8 in mount & \\
         \cline{1-5}
         Thermal cycling & -70 $^{\circ}$C - +70 $^{\circ}$C & 8 cycles with 1 hour per plateau & 42 hours & S1 in mount \par S8 in mount & \\
         \hline

    \end{tabular}
    \caption{Overview of the environmental tests executed within the SUPPPPRESS-project. Item specification glossary: FS - fused silica substrate, 1AR - anti-reflection coating on one side, 1TR - 1 layer twisted (liquid-crystal) retarder}
    \label{tab:environmental}
\end{table}

\begin{figure}[tbp]
\centering
\includegraphics[width = 0.5\textwidth]{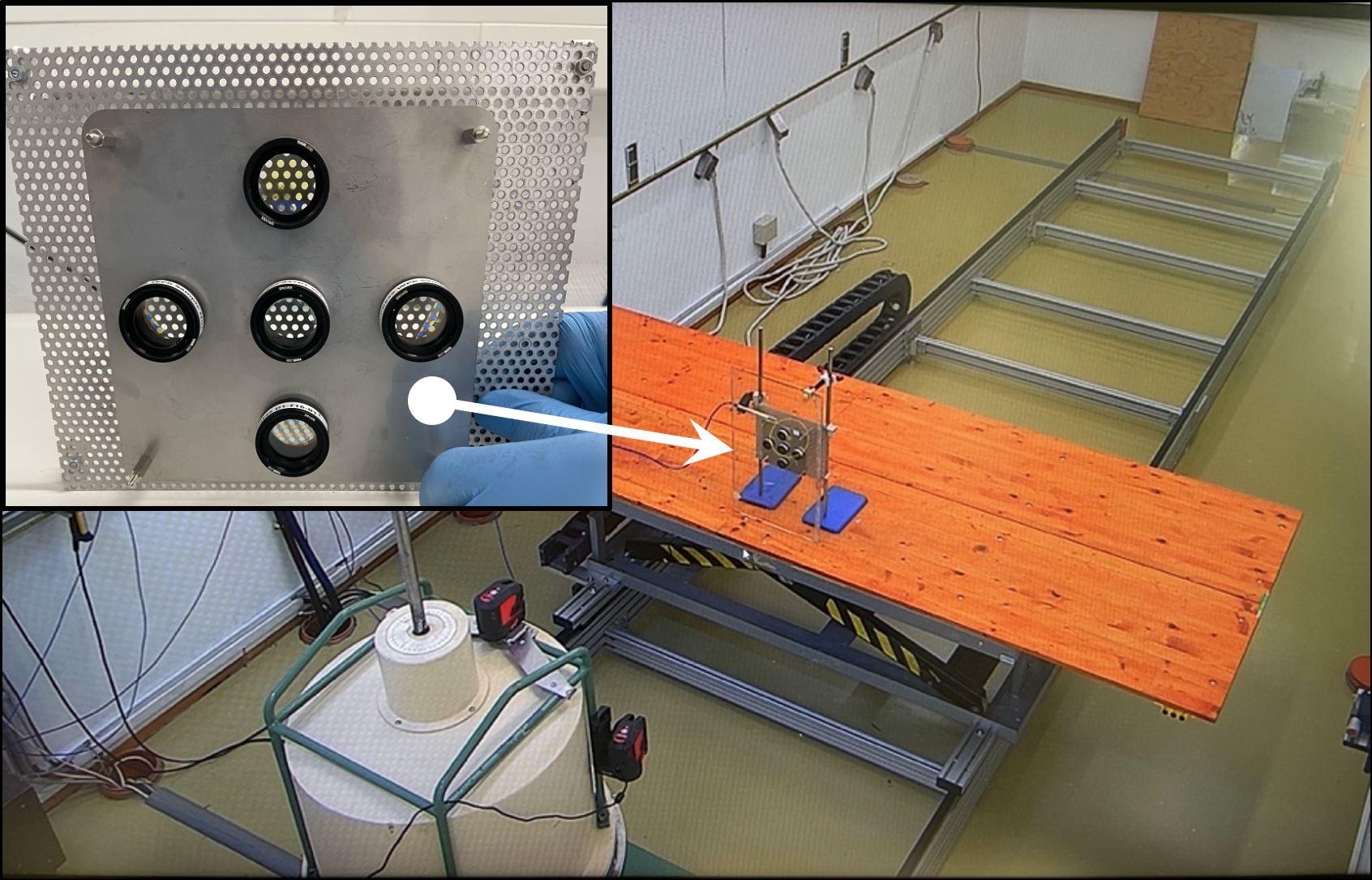}
\caption{Setup at the ESA-ESTEC radiation facility. The Co-60 source is located inside the white vessel at the bottom left. The sample holder (insert top left) is positioned at a translation table at a predetermined distance in order to have the desired fluence at the sample.} 
\label{fig:gamma_setup}
\end{figure}

The $\gamma$-radiation test was executed at the radiation test facility of ESA ESTEC in Noordwijk, the Netherlands. We used a relative low dose rate of 791 rad/hour in order to achieve 33.45 krad of radiation over the course of two days. Five samples were placed in a holder at the distance from the source that corresponds to the desired dose rate, see Fig.~\ref{fig:gamma_setup}.

\begin{figure}[tbp]
\centering
\DIFaddbeginFL \includegraphics[width = \linewidth]{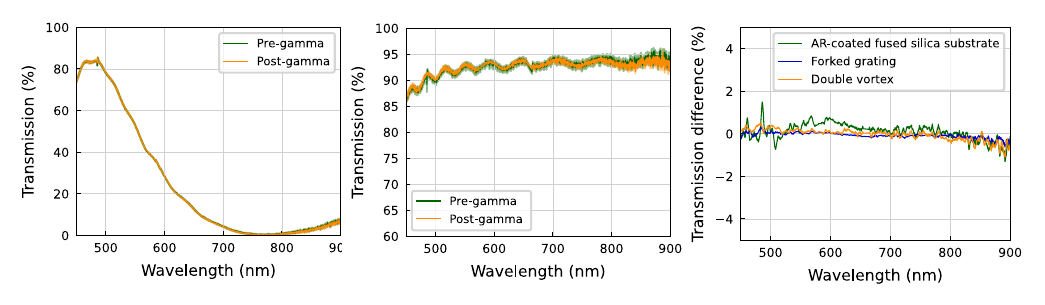}
\DIFaddendFL \caption{Pre- and post-$\gamma$-radiation transmission measurements for an uncapped forked grating (S4, left) and a (capped) double vortex prototype (S5, middle). The right figure shows the transmission difference for these two samples, including a reference sample without liquid-crystal layer (S1).} 
\label{fig:prepostgamma}
\end{figure}

The results of the pre- and post-test transmission measurements are shown in \ref{fig:prepostgamma} for an uncapped (forked) polarization grating on a fused silica substrate and a double vortex prototype, for which the liquid-crystal layers are not exposed to air. The \DIFaddbegin \DIFadd{forked grating shows a minimum of near 0 transmission at the wavelength of $\sim$780 nm as a result of near-perfect half-wave retardance and therefore near-zero on-axis leakage. The double vortex prototype shows a high transmission ($>90\%$) over a broad wavelength range. The fringes are attributed to the properties of the (bulk) liquid crystal layer since reflection losses at the AR-coating, fused silica - liquid crystal and liquid-crystal - NOA61 glue interfaces are expected to be less than 2\%. The }\DIFaddend difference between the pre- and post-$\gamma$-radiation data is smaller than 0.1\% RMS over the full measured spectrum, demonstrating that no lasting transmission effect has been observed after $\gamma$-radiation. In polarization microscope images, no change of number of defects has been observed.

The UV-radiation test was executed in-house using the output of an Energetiq Laser Driven Light Source. A sample was placed in front of the bare output fiber such that a circular area of $\approx$ 13.5 mm diameter was irradiated. The circular area was tuned such that the integrated fluence for wavelengths between 200-400 nm of the specified \DIFdelbegin \DIFdel{LDLS }\DIFdelend \DIFaddbegin \DIFadd{Laser-driven light source (LDLS) }\DIFaddend spectrum roughly matches the integrated solar fluence. Since calibration measurements of the LDLS output showed an enhanced fluence between 380-420 nm and a reduced fluence below 380 nm, compared to the specified LDLS spectrum, the irradiation time was increased from the required minimum of 10 hours to 45 hours for samples 4, 5 and 9 and to 20(24) hours for sample 6(7).

\begin{figure}[tbp]
\centering
\DIFdelbeginFL 
\DIFdelendFL \DIFaddbeginFL \includegraphics[width = \linewidth]{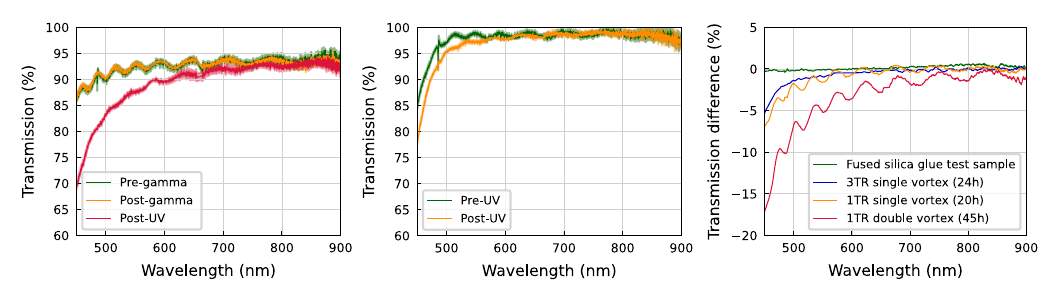}
\DIFaddendFL \caption{Pre- and post-UV-radiation transmission measurements for a 1TR double vortex prototype (S5, left) and a (capped) 1TR single vortex (S6, middle). The right figure shows the transmission difference for these two samples, including a (capped) 3TR single vortex (S7) and a reference glue test sample without liquid-crystal layer (S8).} 
\label{fig:prepostuv}
\end{figure}

The results of the pre- and post-test transmission measurements are shown in \DIFaddbegin \DIFadd{Fig.~}\DIFaddend \ref{fig:prepostuv} for a double vortex, two single vortices (based on a 1TR and 3TR liquid-crystal recipe), and a glue test reference sample without liquid-crystal layer. In all cases the liquid-crystal layer is not exposed to air. A clear difference between the pre- and post-UV-radiation data is observed for the samples with liquid-crystal layer. First, the transmission is reduced with the strongest reduction at the shortest wavelengths, corresponding to about -0.35\% per hour at 450 nm for the 1TR samples and slightly less for the 3TR sample. Second, the fringes observed in the pre-UV-radiation transmission profile of the 1TR-samples are reduced in amplitude and slightly shifted in spectral position. This observation requires further analysis \DIFdelbegin \DIFdel{in order to assess if/}\DIFdelend \DIFaddbegin \DIFadd{to assess whether and }\DIFaddend how this is related to the liquid-crystal layer, since this is not observed for the 3TR sample.

We note that both the liquid-crystal materials and the NOA61 glue require UV-curing in order to properly solidify, and therefore both strongly absorb in the UV: the curing wavelength used is 355 nm for the liquid crystals and 365 nm for the NOA61 glue. This could introduce color centers similar to what is commonly known for optical crystals and glasses under UV-radiation, which \DIFdelbegin \DIFdel{lead }\DIFdelend \DIFaddbegin \DIFadd{leads }\DIFaddend to additional absorption extending into the visible spectrum. Pre- and post UV-test polarization microscope images however did not show a change in defect appearance nor in the liquid-crystal pattern. A quantitative analysis of local retardance changes (if any) after UV-radiation will provide an upper limit for the UV-sensitivity of the birefringence of the liquid-crystal layer. \DIFaddbegin \DIFadd{A final demonstration that the performance at a high-contrast testbed is not impacted by UV-radiation will truly qualify these components for use in space or indicate how much UV-mitigation needs to be applied. Although initially planned within the SUPPPPRESS project, this could not be executed due to schedule constraints.
}\DIFaddend 

\begin{figure}[tbp]
\centering
\DIFdelbeginFL 
\DIFdelendFL \DIFaddbeginFL \includegraphics[width = \linewidth]{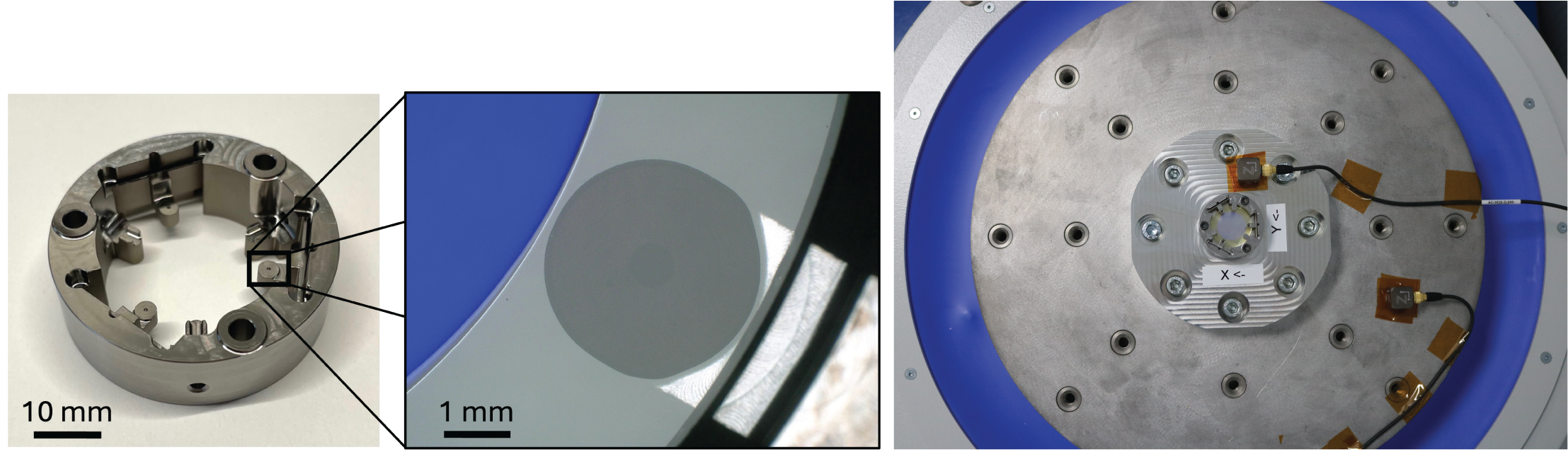}
\DIFaddendFL \caption{Left) The (empty) Ti-structure for mounting patterned liquid-crystal assemblies. Middle) A reflection microscope image of the area around the glue pad of a AR-coated fused silica substrate bonded into the Ti-structure. The dark grey bond spot is clearly visible at the uncoated outer area of the substrate while not touching the (blue) AR-coating. Right) The Ti-mount with a dummy sample consisting of three fused silica substrates of which 2 \DIFdelbeginFL \DIFdelFL{(TBC) }\DIFdelendFL have a patterned liquid-crystal layer mounted to the shaker for out-of-plane testing.} 
\label{fig:ti-mount}
\end{figure}

The mounting concept under test is shown in \DIFaddbegin \DIFadd{Fig.~}\DIFaddend \ref{fig:ti-mount}. It is a single titanium component produced by milling and electrical discharge machining with three glue pads to which the optical component is glued, channels for glue insertion, and features for alignment. Bonding is performed using \DIFdelbegin \DIFdel{space qualified }\DIFdelend \DIFaddbegin \DIFadd{space-qualified }\DIFaddend Scotch EC-2216 glue at a coating-free area of the fused silica substrate. Therefore, the AR-coated first and last substrates have a 3.5 mm coating-free outer rim. This resulted in glue spot sizes of about 3 mm, see \DIFdelbegin \DIFdel{Figure }\DIFdelend \DIFaddbegin \DIFadd{Fig. }\DIFaddend \ref{fig:ti-mount}-middle. For the vibration test a dummy sample consisting of three fused silica substrates of which 2 have a patterned (polarization grating) liquid-crystal layer was assembled and glued into the mount.
The full assembly was mounted on an interface plate and attached to the Tira Vibration Test System TV 56280/LS-340 in both out-of-plane and in-plane configurations. Both sine and random vibration sweeps were executed for the three axis at the levels listed in Table \ref{tab:environmental}\DIFdelbegin \DIFdel{, with the results plotted for the out-of-plane test in Figure \ref{fig:ti-mount}.
}\DIFdelend \DIFaddbegin \DIFadd{.
}\DIFaddend No detachment occurred and no visible degradation at the glue interfaces was observed in microscope observations.

Next, the assembly was put in a ESPEC ARG-0220 thermal cycling chamber and exposed to eight thermal cycles from -70$^{\circ}$C to +70$^{\circ}$C in a nitrogen atmosphere. Also after this test, no detachment occurred, and no visible degradation at the glue interfaces was observed in microscope observations, consistent with an earlier thermal test using a single bare fused silica substrate glued to another instance of the titanium mount. In addition, a separate \DIFdelbegin \DIFdel{double grating }\DIFdelend \DIFaddbegin \DIFadd{double-grating }\DIFaddend assembly participated in the thermal cycling program, also without detachment or visible degradation, confirming that also the liquid crystal - NOA61 - liquid crystal interface is capable of withstanding this thermal range. This is consistent with similar coronagraphic components \DIFaddbegin \DIFadd{that }\DIFaddend are already operational in cryovacuum environments at ground-based telescopes.\cite{Doelman:21AppOpt}

\DIFdelbegin 
{
\DIFdelFL{Test profiles of the sine (top left) and random (bottom left) out-of-plane vibrations and the thermal cycling profile of the Ti-mount with bonded AR-coated fused silica substrate (right).}} 
\DIFdelend 

\DIFaddbegin \DIFadd{These results validate the patterned liquid-crystal technology for coronagraphic components in missions such as HWO, Lazuli\mbox{
\cite{2026arXiv260102556R}}\hskip0pt
, and PICTURE-D\mbox{
\cite{2023SPIE12680E..0FM}}\hskip0pt
. In addition, they strengthen confidence in applying the same liquid-crystal technology to a broader range of functionalities in future space instrumentation. The TOLIMAN\mbox{
\cite{2024SPIE13100E..1RL} }\hskip0pt
mission considers the same technology and assembly processes for the manufacturing of the large diffractive element to enable differential astrometry. Beyond diffractive optics, broadband patterned retarders can also be applied in polarimetric instrumentation. A spatially varying half-wave or quarter-wave retarder can be implemented for snapshot spectropolarimetry with polarization modulation perpendicular to the spectral dimension\mbox{
\cite{2019PASP..131g5002S, 2019SPIE11132E..0AS}}\hskip0pt
. Based on this approach, an updated version of the LOUPE instrument\mbox{
\cite{2021RSPTA.37990577K} }\hskip0pt
is being developed for full-Stokes spectropolarimetry of the Earth-as-an-exoplanet from the Moon, to provide benchmark data of a living planet for observations of biosignatures with HWO and other future space telescopes.}\\
\DIFadd{The multi-layer liquid-crystal recipes can also be modified for a rotating (unpatterned) wave plate that serves as a modulator for a more classical implementation of a polarimeter.
Instead of optimizing the retardance of such a wave plate to be achromatic, one can also create a recipe that provides optimal full-Stokes polarization modulation at all wavelengths within a very broad wavelength range (``polychromatic'' modulation\mbox{
\cite{2012SPIE.8446E..25S, 2015SPIE.9613E..0GS}}\hskip0pt
).
Such a polychromatic polarization modulator based on multi-layer liquid-crystal polymers is being considered for the vis-NIR arms of the Pollux\mbox{
\cite{2025SPIE13699E..28N} }\hskip0pt
UV-vis-NIR spectropolarimeter that is being proposed for HWO.
}

\DIFaddend \section{Conclusions \& outlook}
This paper reports on the progress achieved within the ESA-funded SUPPPPRESS project, which aims to prototype and test liquid-crystal vector vortex coronagraphs (VVCs) with reduced polarization leakage. We have demonstrated improved manufacturing capabilities at \DIFdelbegin \DIFdel{Colorlink }\DIFdelend \DIFaddbegin \DIFadd{ColorLink }\DIFaddend Japan and described the metrology methods used to characterize the fabricated masks. While we are continually improving the manufacturing quality, the project has \DIFdelbegin \DIFdel{already }\DIFdelend \DIFaddbegin \DIFadd{currently }\DIFaddend achieved the following milestones:

\begin{itemize}
    \item Central singularities with diameters of $\sim 2~\mu$m and $\sim 6~\mu$m for charge-2 and charge-6 VVCs, respectively.
    \item Vortex patterns with an azimuthal root-mean-square fast-axis error of $<1^\circ$.
    \item An average polarization leakage of $3 \times 10^{-4}$ over a 10\% bandwidth and $8 \times 10^{-4}$ over a 20\% bandwidth.
    \item Polarization gratings with \DIFdelbegin \DIFdel{no detectable diffraction into }\DIFdelend \DIFaddbegin \DIFadd{diffracted relative intensities of $<10^{-6}$ in }\DIFaddend half-orders \DIFdelbegin \DIFdel{and a relative intensity of }\DIFdelend \DIFaddbegin \DIFadd{and }\DIFaddend $<10^{-3}$ in other unwanted diffraction orders.
    \item Significantly improved defect control and manufacturing yield.
\end{itemize}

In addition, we have developed and implemented tooling for the assembly of multi-component liquid-crystal mask assemblies, including a \DIFdelbegin \DIFdel{double-grating VVC (dgVVC ) }\DIFdelend \DIFaddbegin \DIFadd{dgVVC }\DIFaddend prototype. This enabled an azimuthal clocking accuracy better than $0.1^\circ$. Far-field diffraction measurements of the assembled dgVVC are in good agreement with simulations and do not reveal any assembly-related issues. High-contrast testing at the THD2 bench resulted in an average \DIFdelbegin \DIFdel{monochromatic }\DIFdelend \DIFaddbegin \DIFadd{1\% bandwidth }\DIFaddend contrast of $2 \times 10^{-8}$ between 3 and $10~\lambda/D$ for the dgVVC prototype with a circularly polarized input beam. Broadband measurements yielded average contrasts of $6 \times 10^{-8}$ and $1 \times 10^{-7}$ for bandwidths of 10\% and 20\%, respectively.

\DIFdelbegin \DIFdel{Future work will focus on further reducing the fast-axis error }\DIFdelend \DIFaddbegin \DIFadd{Even with the improved manufacturing quality, further development is required in order to achieve the ambitious $10^{-10}$ raw contrast level set by HWO: First, we will need to further reduce the patterning error of both the vortex as well as the gratings. Decreasing the intensities in higher-order diffraction orders due to writing errors is especially important at the F-number that HWO will likely have. We plan to improve the patterning quality }\DIFaddend through improved calibration procedures and operation of the direct-write machine in a stabilized environment. In parallel, \DIFdelbegin \DIFdel{efforts will continue to enhance manufacturing yield and broadband performance. Planned }\DIFdelend \DIFaddbegin \DIFadd{we aim to further improve the broadband leakage performance of the manufactured components. Furthermore, assemblies with a central opaque dot will be tested to mitigate the impact of the central singularity. On the testbed side, we aim to test the components in dual-polarization mode. This requires the implementation and testing of wavefront control algorithms that can handle both polarization states simultaneously. In addition, future planned }\DIFaddend activities include the assembly and testing of triple-grating VVCs\DIFdelbegin \DIFdel{, as well as testing mgVVCs in dual-polarization mode. Additionally, we }\DIFdelend \DIFaddbegin \DIFadd{. We also }\DIFaddend aim to extend component development toward shorter wavelengths in the blue as well as toward the near-infrared. Following validation of these improvements, the components will need to be tested in vacuum to assess \DIFdelbegin \DIFdel{the }\DIFdelend \DIFaddbegin \DIFadd{their }\DIFaddend ultimate contrast limits. In the case of a segmented aperture, the same technology can be used to manufacture geometric phase apodizers for the VVC, which are achromatic and can be optimized for both circular polarization states simultaneously. \cite{10.1117/12.3064653}\\

\DIFdelbegin \DIFdel{With }\DIFdelend \DIFaddbegin \DIFadd{Finally, with }\DIFaddend the basic environmental tests performed on compound optics with patterned liquid-crystal polymers, the avenue towards full space qualification is clear, with no major issues expected. 
Note that components based on switching nematic liquid crystals (that are still actually ``liquid'' whereas the liquid-crystal polymers \DIFdelbegin \DIFdel{discussing }\DIFdelend \DIFaddbegin \DIFadd{discussed }\DIFaddend in this paper are \DIFdelbegin \DIFdel{solifified }\DIFdelend \DIFaddbegin \DIFadd{solidified }\DIFaddend through UV-curing) have been fully qualified and are currently operational onboard Solar Orbiter\cite{2024SPIE13050E..0HA}.
The main issue observed in our environmental tests is the reduction of transmission\DIFdelbegin \DIFdel{(}\DIFdelend \DIFaddbegin \DIFadd{, }\DIFaddend particularly in the blue \DIFdelbegin \DIFdel{) }\DIFdelend \DIFaddbegin \DIFadd{end of the visible spectrum, }\DIFaddend upon strong UV illumination.
This effect is to be expected for a material that is strongly absorbent in the UV, and this effect may be irrelevant for low in-flight doses of UV irradiation, or easily mitigated through a UV-blocking filter/glass.
In any case, the main properties for these patterned \DIFdelbegin \DIFdel{retardance }\DIFdelend \DIFaddbegin \DIFadd{retarders }\DIFaddend (pattern accuracy and retardance) appear unaffected, although further analysis is required to provide an ultimate upper limit on such degradations.\\
\DIFdelbegin \DIFdel{These results also instil confidence for the application of the same liquid-crystal technology in other future space missions.
The TOLIMAN\mbox{
\cite{2024SPIE13100E..1RL} }\hskip0pt
mission considers the same technology and production/assembly processes for the manufacturing of the large diffractive element to enable differential astrometry.
In addition to diffractive elements, the broadband patterned retarders can also be applied to polarimetric instrumentation.
A space-varying half-wave or quarter-wave retarder can be implemented for snapshot spectropolarimetry with polarization modulation perpendicular to the spectral dimension\mbox{
\cite{2019PASP..131g5002S, 2019SPIE11132E..0AS}}\hskip0pt
.
Based on this approach, an update version of the LOUPE instrument\mbox{
\cite{2021RSPTA.37990577K} }\hskip0pt
is being developed for full-Stokes spectropolarimetry of the Earth-as-an-exoplanet from Moon, to provide benchmark data of a living planet for  observations of biosignatures with HWO and other future space telescopes.}
\DIFdel{The multi-layer liquid-crystal recipes can also be modified for a rotating (unpatterned) wave plate that serves as a modulator for a more classical implementation of a polarimeter.
Instead of optimizing the retardance of such a wave plate to be achromatic, one can also create a recipe that provides optimal full-Stokes polarization modulation at all wavelengths within a very broad wavelength range (``polychromatic'' modulation\mbox{
\cite{2012SPIE.8446E..25S, 2015SPIE.9613E..0GS}}\hskip0pt
).
Such a polychromatic polarization modulator based on multi-layer liquid-crystal polymers is being considered for the vis-NIR arms of the Pollux\mbox{
\cite{2025SPIE13699E..28N} }\hskip0pt
UV-vis-NIR spectropolarimeter that is being proposed for HWO.
}\DIFdelend

\subsection*{Disclosures}
The authors declare that there are no financial interests, commercial affiliations, or other potential conflicts of interest that could have influenced the objectivity of this research or the writing of this paper.

\subsection* {Code, Data, and Materials Availability} 
The data and code used in this work are available upon reasonable request to the authors.

\subsection* {Acknowledgments}
This work was supported by the European Space Agency (ESA) under the tender number TDE-TEC-MOO AO/1-11613/23/NL/AR. Minor language and grammar clean-up was done using ChatGPT\DIFaddbegin \DIFadd{. We thank the anonymous referees for their constructive reports which have significantly improved the quality of this work}\DIFaddend .


\bibliography{report}   
\bibliographystyle{spiejour}   


\vspace{2ex}\noindent\textbf{} 

\vspace{1ex}


\end{document}